\documentclass[showpacs,preprintnumbers,amsmath,amssymb,APSl,prd,nofootinbib,superscriptaddress, twocolumn]{revtex4-2}

\usepackage{amsmath}
\usepackage{amssymb}
\usepackage{amsfonts}
\usepackage{bbold}
\usepackage{graphicx,bm}
\usepackage{dcolumn}
\usepackage[colorlinks=true]{hyperref}
\usepackage{epsf}
\usepackage{xfrac}
\usepackage{enumerate}
\usepackage{hhline}
\usepackage{array}
\usepackage{tabularx}
\usepackage{color}
\usepackage{float}
\usepackage{subcaption}
\usepackage{orcidlink}

\newcommand{\be}{\begin{equation}}
\newcommand{\ee}{\end{equation}}
\newcommand{\bea}{\begin{eqnarray}}
\newcommand{\eea}{\end{eqnarray}}
\newcommand{\beaa}{\begin{eqnarray*}}
\newcommand{\eeaa}{\end{eqnarray*}}

\def\be{\begin{equation}}
\def\ee{\end{equation}}
\def\bea{\begin{eqnarray}}
\def\eea{\end{eqnarray}}

\begin{document}

\title{Analyzing Quasi-normal modes spectrum in asymptotically de Sitter black bounces and their optical appearance}
\author{G. Alencar
\orcidlink{0000-0002-3020-4501}
}
\email{geova@fisica.ufc.br}
\affiliation{Department of Physics, Universidade Federal do Cear\'a (UFC), Campus do Pici, Fortaleza - CE, C.P. 6030, 60455-760 - Brazil}

\author{Albert Duran-Cabac\'es
\orcidlink{0000-0002-2646-7763}
} 
\email{albert.duran22@uva.es}
\affiliation{Department of Theoretical Physics, Atomic and Optics, and Laboratory for Disruptive Interdisciplinary Science (LaDIS), Campus Miguel Delibes, \\ University of Valladolid UVA, Paseo Bel\'en, 7,
47011 - Valladolid, Spain}

\author{A. Lima 
\orcidlink{}}
\email{arthur.lima@fisica.ufc.br}
\affiliation{Department of Physics, Universidade Federal do Cear\'a (UFC), Campus do Pici, Fortaleza - CE, C.P. 6030, 60455-760 - Brazil}

\author{\'Angel Rinc{\'o}n 
        \orcidlink{0000-0001-8069-9162}} 
        \email{angel.rincon@physics.slu.cz}
        \affiliation{Departamento de F{\'i}sica, Universidad del B{\'i}o-B{\'i}o, Casilla 5-C, Concepci{\'o}n, Chile.}
        \affiliation{Research Centre for Theoretical Physics and Astrophysics, Institute of Physics, Silesian University in Opava, Bezrucovo n\'{a}m. 13, 74601 Opava, Czech Republic.}

\author{Diego S\'aez-Chill\'on G\'omez
        \orcidlink{0000-0002-0714-2680}}
\email{diego.saez@uva.es} 
\affiliation{Department of Theoretical Physics, Atomic and Optics, and Laboratory for Disruptive Interdisciplinary Science (LaDIS), Campus Miguel Delibes, \\ University of Valladolid UVA, Paseo Bel\'en, 7,
47011 - Valladolid, Spain}
\affiliation{Department of Physics, Universidade Federal do Cear\'a (UFC), Campus do Pici, Fortaleza - CE, C.P. 6030, 60455-760 - Brazil}

\begin{abstract}

This work explores asymptotically de Sitter black-bounce spacetimes, obtained as a regular extension of the Schwarzschild–de Sitter geometry via the Simpson–Visser prescription. The resulting family of solutions provides a smooth interpolation between regular black holes, extremal configurations, and traversable wormholes within a single geometrical framework. 
In particular, we verify the regularity of the spacetime by showing that (some of) curvature invariants, including the Kretschmann and Ricci scalars, remain finite at the origin for any non-vanishing value of the regularization parameter $a$, ensuring geodesic completeness. An analysis of the effective stress-energy tensor reveals a violation of the null energy condition, which can be consistently interpreted in terms of a composite source involving a phantom scalar field coupled to nonlinear electrodynamics.
Subsequently, we investigate the observational signatures of these geometries by studying: 
 i) null geodesic propagation and 
ii) constructing the associated optical appearance. 
In particular, we determine the shadow radius and characterize the structure of photon rings through numerical ray-tracing procedure. 
Finally, we compute the spectrum of axial gravitational perturbations using a pseudospectral approach, obtaining the quasinormal mode frequencies and their overtones.
Our results show a non-trivial interplay between the cosmological constant and the regularization scale $a$, which imprints itself on both the light-ring structure and the quasinormal spectrum. These features suggest potential observational discriminants that could distinguish these regular spacetimes from their singular counterparts in future gravitational-wave and high-resolution electromagnetic observations.

\end{abstract}

\maketitle


\section{Introduction}

In the context of strong gravity, black holes (BH hereafter) represent one of the most fascinating predictions of General Relativity (GR). There are several reasons, but roughly speaking, BHs are ideal ``laboratories'' to test i) strong-field gravity, ii) quantum effects, iii) the interplay between gravity and cosmology, among other interesting effects (see \cite{Grumiller:2022qhx,Chandrasekhar:1985kt,Carballo-Rubio:2025fnc} and references therein).
From the plethora of black holes theoretically predicted, the Kerr solution \cite{Kerr:1963ud} has a special spot: its metric describes rotating black holes in asymptotically flat spacetime. However, when we consider realistic astrophysical environments a positive cosmological constant $\Lambda > 0$ is required, in general, leading to an asymptotically de Sitter (dS) spacetime \cite{Henneaux1985}. Thus, the Schwarzschild-de Sitter (SdS) and Kerr-de Sitter solutions exemplify this class, featuring both a black hole event horizon and a cosmological horizon.
In recent years, regular (or more precisely, non-singular) black hole solutions and ``black bounce'' spacetimes have gained considerable attention as alternatives that ``remove'' singularities while preserving key phenomenological features of black holes (for a review, see \cite{Lan:2023cvz}).

The canonical example of the implementation of the ``bounce'' idea is the Simpson-Visser black-bounce metric \cite{Simpson:2018tsi}, where a minimal modification of the Schwarzschild geometry is introduced by replacing the radial coordinate $r \to \sqrt{r^2 + a^2}$ (with $a$ being a regularization parameter). Furthermore, such a parameter interpolates between a regular black hole and a traversable wormhole. As a consequence, much attention has been drawn to black bounces/regular black holes in recent years, where some other solutions have been proposed and the corresponding sources have been obtained \cite{Franzin:2021vnj, Simpson:2019cer, Crispim:2024yjz, Crispim:2025cql, Alencar:2025nik, Bronnikov:2005gm, Ayon-Beato:1998hmi, Bronnikov:2021uta, Alencar:2025jvl, Rodrigues:2023vtm, Carballo-Rubio:2019fnb, Torres:2022twv, Bueno:2024dgm}. Nevertheless, most of these spacetime metrics are constructed to be asymptotically flat. Then, extensions of such models to asymptotically de Sitter spacetimes provide a rich and interesting framework for exploring regular geometries in cosmological backgrounds. These ``asymptotically de Sitter black bounces'' combine the bounce regularization with a positive cosmological constant, yielding spacetimes that are regular at the core and exhibit both black hole and cosmological horizons depending on the parameter values.

Now that we have established that the study of asymptotically de Sitter black bounces deserves considerable attention, it becomes essential to mention which properties could be more interesting to investigate. Given that we are in the golden age of multimessenger astronomy, we will focus on two of them: the corresponding quasinormal modes (QNMs) and shadows (see several examples  in \cite{Tsupko:2018apb,Frolov:1998wf,Rincon:2018sgd,Contreras:2019cmf,Rincon:2020cos,Panotopoulos:2019qjk,Rincon:2018ktz,Panotopoulos:2020mii,Rincon:2026ooi,Koch:2025gaw,Diaz-Guerra:2025jip,Gonzalez:2022ote,Alencar:2025yyl,Olmo:2025ctf} and references therein).
Firstly, the QNMs are usually defined as the characteristic damped oscillations of perturbed black holes, which contain crucial information about: i) their stability, ii) relaxation timescales, and iii) response to external perturbations. Irrespective of the internal details, the QNM spectrum is determined by solving the linearized perturbation equations with appropriate boundary conditions (ingoing at the event horizon and outgoing at spatial/cosmological infinity) as can be consulted in  \cite{Berti:2009kk,Bolokhov:2025uxz,Nollert:1999ji,Konoplya:2019hlu,Konoplya:2004ip} and references therein. Analysis of the QNM spectrum can lead to tests of the Kerr family of solutions, as widely pointed out in the literature \cite{Carullo:2018sfu,Isi:2019aib,Cardoso:2016oxy,Cardoso:2017cqb,LIGOScientific:2020tif}. Actually, the observations of the event GW250114 have achieved the calculation of the first overtone in the QNM spectrum, which is used to test any deviations from GR canonical solutions \cite{LIGOScientific:2025wao}. Despite that every evidences conclude that the final object involved in the GW250114 merge is a Kerr black hole, the quick increasing number of events detected and expected for the next few years make fundamental any study concerning possible imprints of deviations from Kerr-like spacetimes on the QNM spectrum.
For asymptotically dS spacetimes, the presence of the cosmological horizon modifies the boundary conditions and the resulting spectrum compared to asymptotically flat cases.
For black-bounce geometries, QNMs have been studied in various contexts \cite{Duran-Cabaces:2025sly,Santos:2025xbk,Malik:2024qsz}, revealing deviations from the Schwarzschild/Kerr spectra that depend on the bounce parameter. Analyzing the QNM spectrum of asymptotically dS black bounces is essential for understanding their stability under perturbations and potential gravitational wave signatures detectable by current and future observatories.

A closely related computation useful to supplement the dynamical information encoded into QNMs is the optical appearance of black holes (particularly their shadows), providing a powerful and also non-trivial probe of the strong-field regime.
It is well-known that the black hole shadow can be ``defined'' as the dark region in the sky formed by the capture of light rays by the photon sphere, surrounded by a bright ring of gravitationally lensed photons \cite{Bronzwaer:2021lzo,Gralla:2019xty,Ovgun:2024zmt,Pulice:2023dqw}. In asymptotically dS spacetimes, the shadow is altered by both the black hole parameters and the cosmological constant, with observers potentially located between the horizons or at different cosmic times.
In addition, the study of shadows is now more attractive given the recent Event Horizon Telescope (EHT) observations of M87* and Sgr A* \cite{EventHorizonTelescope:2019dse,EventHorizonTelescope:2022xqj}: such observations have established stringent constraints on black hole parameters and deviations from the Kerr metric \cite{EventHorizonTelescope:2021dqv}. Despite every evidence point to a Kerr black hole for the central objects located at M87* and Sgr A*, the beginning of the golden era of very-long baseline interferometry, which expects to extend the image reconstruction procedure to an increasing number of ultra-compact objects, has pushed forward possible smoking guns for testing the nature of objects that deviate from the Kerr paradigm \cite{Addazi:2021xuf,AlvesBatista:2023wqm,Cardoso:2019rvt,Staelens:2023jgr,Guerrero:2021ues,Kocherlakota:2023qgo,Chael:2021rjo,Vincent:2022fwj,Olmo:2023lil,daSilva:2023jxa,DeMartino:2023ovj,Olmo:2021piq,Chen:2024ibc}. Extending shadow calculations to regular dS black bounces allows for testing the viability of singularity-free models against observational data and exploring how the bounce parameter and $\Lambda$ affect the shadow radius, distortion, and photon ring structure when considering a spacetime that is not asymptotically flat \cite{Alencar:2025yyl}. 

In addition, the idea of combining both sources of data, the light rings' structure of the images and the QNMs spectrum, which actually show a kind of correspondence up to some limit \cite{Cardoso:2008bp}, has become very popular over the last years \cite{Stefanov:2010xz,Jusufi:2019ltj,Pedrotti:2024znu,Konoplya:2024lir,Konoplya:2024vuj,Konoplya:2024lch,Pedrotti:2025upg}. Moreover, it gives a new way of computing with great accuracy the QNMs spectrum departing from properties of the images that are much easier to be obtained \cite{Diaz-Guerra:2026hxm}.

In this work, we investigate the spectrum of quasi-normal modes for axial gravitational perturbations in asymptotically de Sitter black-bounce spacetimes. We employ a pseudospectral approach, obtaining the quasinormal mode frequencies and studying their dependence on the bounce parameter, multipole number, and cosmological constant. Additionally, we analyze the optical appearance by performing null geodesic ray-tracing to determine the shadow and photon rings as seen by static observers. Our results reinforce how the regularization parameter $a$ modifies both the ringing modes and the observable shadow, providing distinguishable signatures from singular black holes.
The paper is organized as follows:  

After this brief introduction, we summarize the basic ingredients of the asymptotically de Sitter black bounces in Sect. \eqref{AsSbb}. Particular attention is given to key aspects such as the role of $  \alpha  $ as an effective cosmological constant and its relation to the admissible mass values that yield positive real roots. We also examine the energy conditions in Subsection \eqref{ECfRSdSS}, along with the regularity of the solution arising from the inclusion of nonlinear electrodynamics \eqref{FSfRSsSS}.
Subsequently, in Section \eqref{GaOotSdSBB}, we analyze the geodesic and optical properties of this spacetime. This discussion is divided into three main parts:
i) null geodesic analysis, where we define the conserved quantities, the effective potential, and the critical impact parameter, leading to the equation that governs the black hole shadow;
ii) the transfer function; and
iii) the optical appearance. The latter two subsections explain the meaning of the transfer function (which maps an observer’s viewing screen to the physical locations where light rays intersect the emitting accretion disk) and present the specific intensity profile used to describe the disk’s emission.
Finally, we introduce the basic theory of quasinormal modes (QNMs) and the pseudospectral method employed to compute the corresponding quasinormal frequencies, together with the numerical results, in Section \eqref{QNMs}. The discussion of the results and the conclusions are presented in Sect. \eqref{Conc}.
We use the mostly positive signature $(-, +, +, +)$ and set $c = 1$ throughout the paper.

\section{Asymptotically de Sitter black bounces}\label{AsSbb}

To construct a regular metric for an asymptotically de Sitter spacetime, we can start from a known singular metric that has the same asymptotic behavior and perform some modification in the definition of the metric that renders it regular. A well-known way to perform this regularization is through the transformation $r^2\rightarrow r^2+a^2$ in the radial coordinate, where $a$ is a regularization parameter. This type of regularization is known as ``black bounce'' and can be applied to several metrics. The original black bounce metric was obtained by Simpson and Visser from the Schwarzschild solution \cite{Simpson:2018tsi}, leading to the following result,
\begin{equation}\label{SVmetric}
ds^2=-A(r)dt^2+\frac{dr^2}{A(r)}+(r^2+a^2)d\Omega^2\ ,
\end{equation}
where
\begin{equation}\label{2}
A(r)=1-\frac{2M}{\sqrt{r^2+a^2}}\ ,
\end{equation}
and $d\Omega^2=d\theta^2+\sin^2\phi$, where  $d\Omega^2$ is the corresponding solid angle. As a result, we obtain a regular solution that interpolates between a regular black hole (when $a<2M$) and a traversable wormhole (when $a>2M$), in addition to generating a wormhole with an extremal throat in the specific case in which $a=2M$. Obviously, the usual Schwarzschild solution is recovered when $a=0$.

\begin{figure}[t!]
\centering
\includegraphics[scale=0.7]{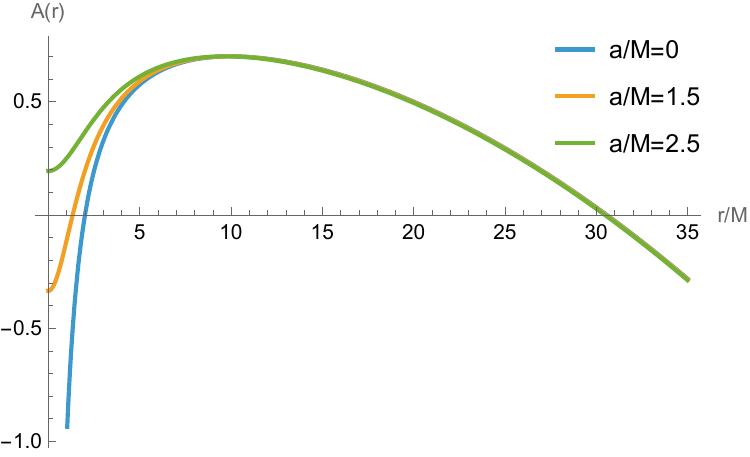}
\caption{The metric function $A(r)$ for different values of the parameter  $a$  of a Schwarzschild-dS regularized black hole.}
\label{Fig1}
\end{figure}

Let us perform the same procedure, but on the singular Schwarzschild–de Sitter metric, which is given by
\begin{equation}\label{3}
ds^2=-A(r)dt^2+\frac{dr^2}{A(r)}+r^2d\Omega^2\ ,
\end{equation}
where
\begin{equation}\label{4}
A(r)=1-\frac{2M}{r}-\alpha^2r^2\ .
\end{equation}
Here, $\alpha^2\equiv \frac{\Lambda}{3}$ with $\Lambda$ being a positive cosmological.
By applying the Simpson–Visser regularization, the line element (\ref{SVmetric}) turns out,
\begin{equation}\label{5}
A(r)=1-\frac{2M}{\sqrt{r^2+a^2}}-\alpha^2(r^2+a^2)\ .
\end{equation}
In Fig.~\ref{Fig1} the function $A(r)$ is depicted for some different values of $a$.

First, let us perform a basic analysis of the causal structure of the metric given in (\ref{5}). To obtain the possible horizons, we must consider null hypersurfaces, for which the normal vectors are null. Since the solution has spherical symmetry, we must solve the expression
\begin{equation}\label{6}
g_{tt}(r=r_{EH})=0\rightarrow A(r=r_{EH})=0\ .
\end{equation}

To simplify our analysis, let us first assume that $a=0$, which corresponds to the usual Schwarzschild–de Sitter solution. Substituting (\ref{5}) into (\ref{6}), the following equation is obtained
\begin{equation}\label{7}
\alpha^2r^3-r+2M=0\ ,
\end{equation}
which is a third-degree algebraic equation. By evaluating the function $A(r)$, one finds that it has a local maximum for $r>0$. This maximum occurs at $r_{max}=\left(\frac{M}{\alpha^2}\right)^{1/3}$, that is, at $A(r_{max})=1-3(M\alpha)^{2/3}$. For the metric to admit some positive real solution, we must assume that $A(r_{max})>0$, since we know that in the limits $r\rightarrow0$ and $r\rightarrow+\infty$, we have $A(r)\rightarrow -\infty$. In other words, the function starts negative near the origin and is also negative in the asymptotic limit; therefore, if the maximum value is also negative, the function never assumes positive values for $r>0$ and thus does not intersect the $r$ axis. Then, the following condition for the metric to have positive real roots yields
\begin{equation}\label{8}
M\alpha<\frac{1}{\sqrt{27}}\ .
\end{equation}

In this case, as shown in figure (\ref{Fig1}), two roots arise, that is, two horizons. The inner horizon corresponds to the event horizon, while the outer one corresponds to the cosmological horizon. For $r<0$, the function is always increasing and tends to $-\infty$ when $r\rightarrow -\infty$, and to $+\infty$ when $r\rightarrow 0$. Therefore, the curve necessarily intersects the $r$ axis at some value of $r<0$, which corresponds to the third real root of equation (\ref{7}).

Assuming that $\alpha<<1$, we can neglect the term $\alpha^2r^2$ for small values of $r$, so that the metric reduces to the Schwarzschild case. Therefore, we can approximate the event horizon as in the asymptotically flat case:
\begin{equation}\label{9}
r_{EH}\approx 2M\ .
\end{equation}

For the asymptotic limit $r>>2M$, we can neglect the term $2M/r$ and approximate the function $A(r)$ as $A(r)\approx 1-\alpha^2r^2$. Hence, the cosmological horizon, in the case in which it is very far from the event horizon, can be approximated by the following
\begin{equation}\label{10}
r_{CH}\approx 1/\alpha\ .
\end{equation}

For the sake of this paper, these limits are used to simplify the  calculations below. Now that we have reviewed these analyzes for the case $a=0$, we can solve the original problem in which $a\neq 0$. In this case, the equation that gives the horizons is
\begin{equation}\label{11}
A(r)=0\rightarrow \alpha^2(r^2+a^2)^{3/2}-\sqrt{r^2+a^2}+2M=0\ .
\end{equation}

The change of the radial coordinate, $x=\sqrt{r^2+a^2}$, can be applied to simplify the calculations. Then, equation (\ref{11}) becomes exactly the same form as (\ref{7}), where $a=0$ is assumed. Then, its roots turn out the same as in that first case. The locations of the horizons in the original radial coordinate are obtained by the relation between the radial coordinates, such that since for the variable $x$ there are two positive solutions corresponding to the event horizon and the cosmological horizon, two possible locations for the horizons in the coordinate $r$ might arise as well,
\begin{eqnarray}
r^2+a^2=x_{EH}^2\rightarrow r_{EH}=\pm \sqrt{x_{EH}^2-a^2}\ ,\label{12}\\ r^2+a^2=x_{CH}^2\rightarrow r_{CH}=\pm \sqrt{x_{CH}^2-a^2}\ .\label{13}
\end{eqnarray}

Let us first evaluate the event horizons. Analogously to the basic black bounce solutions, we see that if $a>x_{EH}$ there are no real event horizons, and thus we obtain a traversable wormhole. For $a<x_{EH}$ we obtain two event horizons with opposite signs, one on the positive side of the radial coordinate and the other one in the negative part, characterizing a bounce between universes. In the extreme case in which $a=x_{EH}$, we obtain an event horizon at the origin, that is, a horizon at the ``throat'', which characterizes a one-way wormhole.

Regarding the second horizon, we see that for the cases in which we have either a black hole or an extremal wormhole, we certainly have $a<x_{CH}$, since $x_{CH}>x_{EH}$. Thus, two cosmological horizons arise, one in each side. However, for the case of a traversable wormhole it is possible to have $a\geq x_{CH}$. In such a case, we have neither event horizons nor cosmological horizons. If $a=x_{CH}$, the throat of the wormhole would lie exactly on the cosmological horizon.

\subsection{Energy Conditions for Regular Schwarzschild de Sitter solution} \label{ECfRSdSS}
To analyze the energy conditions of the solution defined by the metric (\ref{5}), we must first solve the Einstein equations in order to determine the stress–energy tensor that generates this solution, and then associate the components of this tensor with the mass–energy density and the radial and tangential pressures. Once the density and pressures are determined, we can evaluate some important energy conditions, such as the Null Energy Condition (NEC), the Weak Energy Condition (WEC), and the Strong Energy Condition (SEC). The Einstein equation for this metric is given by
\begin{equation}\label{17}
G^{\mu}{}_{\nu}+\delta^{\mu}{}_{\nu}\Lambda=\kappa^2T^{\mu}{}_{\nu}\ ,
\end{equation}
where $\kappa^2=8\pi G$. The nonvanishing components of the Einstein tensor for the regularized Schwarzschild–de Sitter metric are
{\small\begin{eqnarray}
G^t{}_t&=&-3\alpha^2+a^2\frac{\left[\alpha^2(r^2+a^2)^{3/2}+\sqrt{r^2+a^2}-4M\right]}{(r^2+a^2)^{5/2}};\label{18}\\ G^r{}_r&=&\alpha^2\left(\frac{3a^2}{r^2+a^2}-3\right)-\frac{a^2}{(r^2+a^2)^2};\label{19}\\ G^{\theta}{}_{\theta}&=&G^{\phi}{}_{\phi}\nonumber\\&=&-3\alpha^2+a^2\frac{\left[\alpha^2(r^2+a^2)^{3/2}+\sqrt{r^2+a^2}-M\right]}{(r^2+a^2)^{5/2}}\ .\label{20}
\end{eqnarray}}

The stress–energy tensor is defined in terms of the mass–energy density and the radial and tangential pressures. Outside the event horizon (that is, when $A(r)>0$), it simply yields
\begin{equation}\label{21}
T^t{}_t=-\rho;, T^r{}_r=p_{\|};; T^{\theta}{}_{\theta}=T^{\phi}{}_{\phi}=p_{\perp}\ .
\end{equation}

Thus, by substituting the Einstein tensor (\ref{18})-(\ref{21}) in the field equations (\ref{17}), one obtains
\begin{eqnarray}
\rho&=&-a^2\frac{\left[\alpha^2(r^2+a^2)^{3/2}+\sqrt{r^2+a^2}-4M\right]}{\kappa^2(r^2+a^2)^{5/2}};\label{22}\\ p_{\|}&=&a^2\frac{[3\alpha^2(r^2+a^2)^{3/2}-\sqrt{r^2+a^2}]}{\kappa^2(r^2+a^2)^{5/2}};\label{23}\\ p_{\perp}&=&a^2\frac{[\alpha^2(r^2+a^2)^{3/2}+\sqrt{r^2+a^2}-M]}{\kappa^2(r^2+a^2)^{5/2}}\ .\label{24}
\end{eqnarray}

\begin{figure}[t!]
\centering
\includegraphics[scale=0.7]{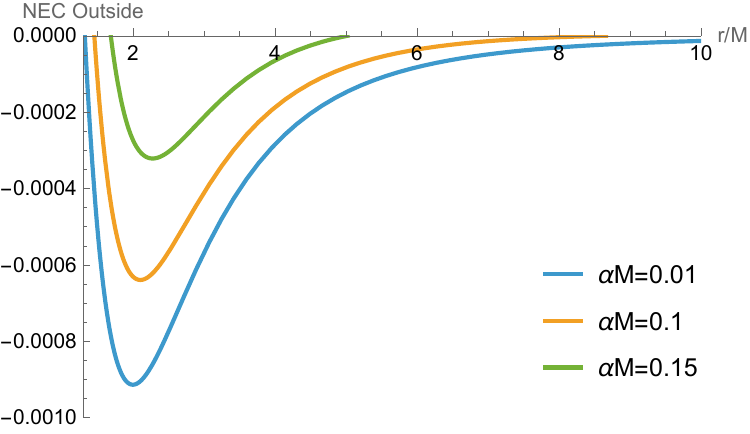}
\includegraphics[scale=0.7]{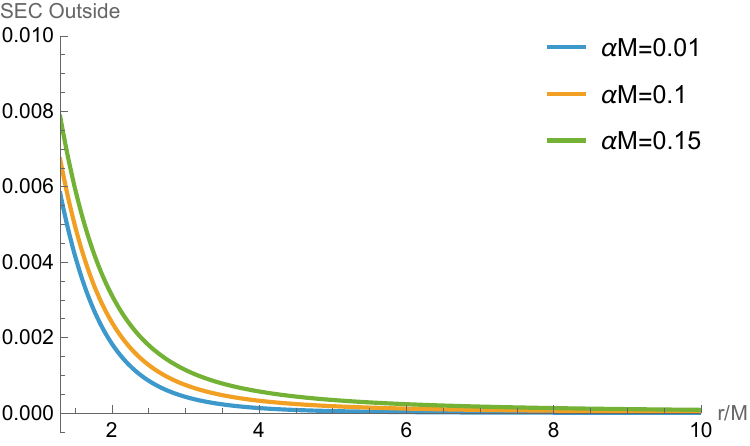}
\caption{The NEC (top panel) and the SEC (bottom panel) outside the event horizon for the regular Schwarzschild-dS spacetime. Some values of $\alpha$ are considered.}
\label{Fig3}
\end{figure}

With these results, the following energy conditions can be easily evaluated.
{\small
\begin{eqnarray}
\text{NEC}:\, &&\rho+p_{\|}=-\frac{2a^2}{\kappa^2(r^2+a^2)^2}A(r)\ ;\label{25}\\ \text{SEC}:\, &&\rho+p_{\|}+2p_{\perp}=2a^2\frac{\left[2\alpha^2(r^2+a^2)^{3/2}+M\right]}{\kappa^2(r^2+a^2)^{5/2}}\ .\label{26}
\end{eqnarray}}
Then, the expression (\ref{25}) reveals that the null energy condition is always violated over the whole spacetime outside the event horizon, since the right-hand side of (\ref{25}) carries an overall negative sign and $A(r)$ is always positive in such a region. On the other hand, the strong energy condition is always satisfied, as the right-hand side of (\ref{26}) is always positive. Figure \ref{Fig3} shows  both conditions for different values of $\alpha$.

The Weak Energy Condition, on the other hand, is associated with the energy density given by equation (\ref{22}). In this case, the behavior of the WEC is not as straightforward, since the term inside the brackets on the right-hand side of (\ref{22}) tends to change sign depending on whether $\alpha^2(r^2+a^2)^{3/2}+\sqrt{r^2+a^2}$ is greater or smaller than $4M$. For $\alpha<<1$, where we may neglect the effect of the term involving $\alpha^2$, and considering that the event horizon is approximately located at $2M$, we have that if $2M<\sqrt{r^2+a^2}<4M$, the expression $\sqrt{r^2+a^2}-4M$ is negative and the energy density is positive, since it depends on an overall negative sign multiplying this expression. For the interval $\sqrt{r^2+a^2}>4M$, the weak energy condition is violated, since $\rho<0$.

\begin{figure}[t!]
\centering
\includegraphics[scale=0.7]{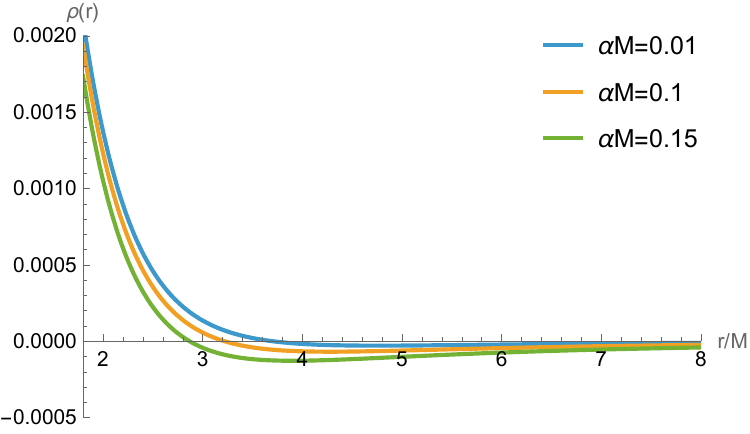}
\caption{The WEC outside the event horizon for some values of $\alpha$ for the regular Schwarzschild-dS metric.}
\label{Fig5}
\end{figure}

With the influence of the cosmological constant, one might have $\alpha^2(r^2+a^2)^{3/2}+\sqrt{r^2+a^2}>4M$ for any $r$ which would imply that the WEC is always violated outside the black hole. However, we have the constraint on $\alpha$ given in (\ref{8}), which limits the possibility of having a complete violation of the WEC. In the case where $M=1$ and $a=1.5$, the maximum value of $\alpha$ is $1/\sqrt{27}\approx 0.19$, and in this situation the weak energy condition is violated, but there exists an initial region in which it is satisfied, similar to the case where $\alpha$ is negligible. This behavior can be better visualized in Fig.~\ref{Fig5}.

For the region inside the event horizon ($A(r)<0$), there is an interchange between the timelike and spacelike characters of the coordinates, which implies the following change in the stress–energy tensor: $T^t{}_t=p_{\|};, T^r{}_r=-\rho$. In this case, the energy density and the radial pressure become
\begin{eqnarray}
\rho&=&a^2\frac{\left[\sqrt{r^2+a^2}-3\alpha^2(r^2+a^2)^{3/2}\right]}{\kappa^2(r^2+a^2)^{5/2}}\ ;\label{27}\\ p_{\|}&=&a^2\frac{\left[\alpha^2(r^2+a^2)^{3/2}+\sqrt{r^2+a^2}-4M\right]}{\kappa^2(r^2+a^2)^{5/2}}\ .\label{28}
\end{eqnarray}

The tangential pressure remains unchanged, since the change in sign of the metric only involves the temporal and radial indices. The Null and Strong Energy Conditions turns out
\begin{eqnarray}
\text{NEC}:\, &&\rho+p_{\|}=\frac{2a^2}{\kappa^2(r^2+a^2)^2}A(r)\ ;\label{29}\\ \text{SEC}:\, &&\rho+p_{\|}+2p_{\perp}=2a^2\frac{\left(2\sqrt{r^2+a^2}-3M\right)}{\kappa^2(r^2+a^2)^{5/2}}\ .\label{30}
\end{eqnarray}
We observe that the null energy condition has the same form as in the region outside the event horizon, but with an overall positive sign. Since $A(r)<0$, the null energy condition is violated again for any $r$ below the event horizon.

\begin{figure}[t!]
\centering
\includegraphics[scale=0.7]{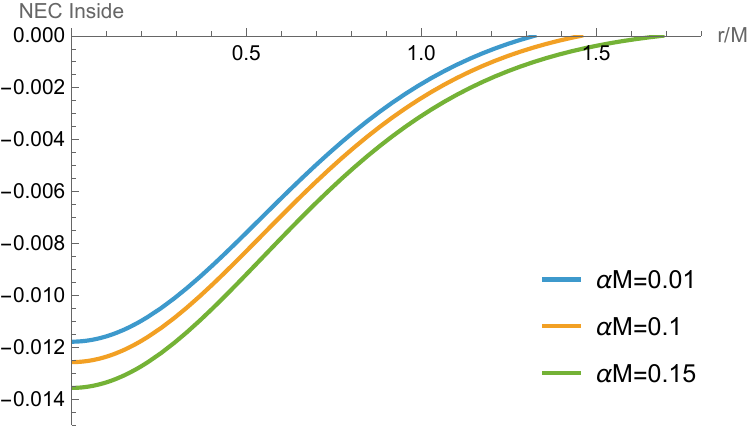}
\includegraphics[scale=0.7]{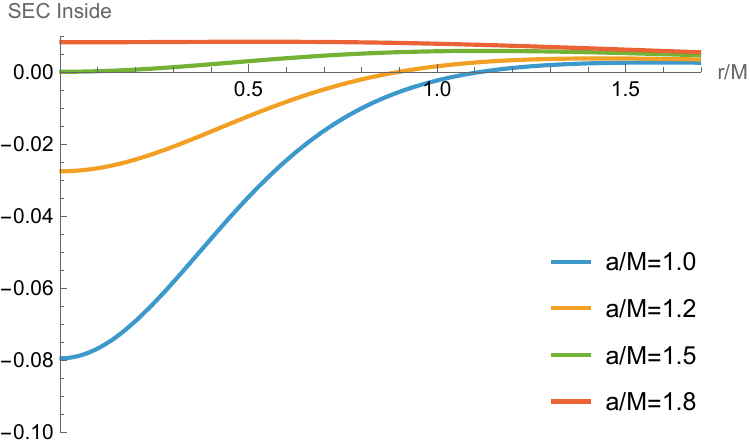}
\caption{The NEC (top) and the SEC (bottom) inside the event horizon for different values of $\alpha$ and $a$, respectively.}
\label{Fig6}
\end{figure}

The strong energy condition exhibits a particular feature, as it does not depend on the cosmological constant. We find that it is violated in the interval $\sqrt{r^2+a^2}<3M/2$, but is satisfied when $\sqrt{r^2+a^2}>3M/2$. By assuming that $r_{EH}\approx 2M$, we see that $3M/2$ may lie within the black hole interior; however, this depends on the parameter $a$, since if $a\geq 1.5M$, then $\sqrt{r^2+a^2}\geq 1.5M$, and consequently there is no violation of the SEC. These behaviors can be observed in Fig.~\ref{Fig6}.

By analyzing expression (\ref{27}) for $\rho$, we can evaluate the WEC. Here, we must analyze the behavior of $1-3\alpha^2(r^2+a^2)$ in order to determine the sign of $\rho$. For the weak energy condition to be satisfied, one must have $\sqrt{r^2+a^2}<\frac{1}{\sqrt{3}\alpha}$, and $\sqrt{r^2+a^2}>\frac{1} {\sqrt{3}\alpha}$ to be violated.

\begin{figure}[H]
\centering
\includegraphics[scale=0.7]{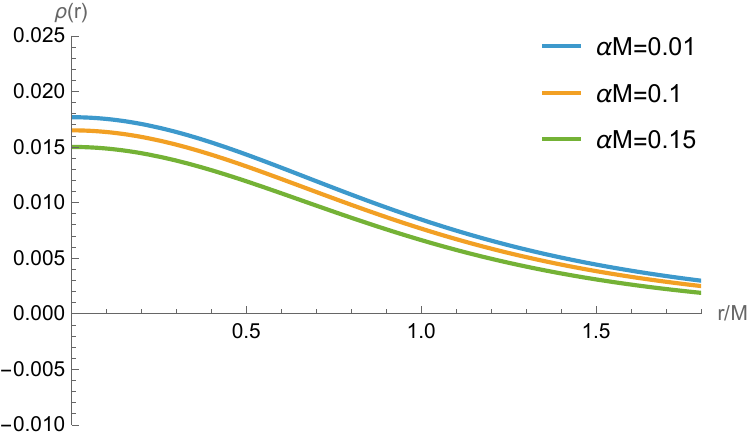}
\caption{The WEC inside the event horizon for some values of $\alpha$ for the metric regular Schwarzschild-dS metric.}
\label{Fig8}
\end{figure}

By assuming $\alpha<<1$, then, $1/\alpha>>1$, and consequently the term $\frac{1}{\sqrt{3}\alpha}$ will certainly be larger than the event horizon. Hence, one always leads to $\sqrt{r^2+a^2}<\frac{1}{\sqrt{3}\alpha}$, that is, the weak energy condition is satisfied. The plot \ref{Fig8} depicts this behavior for the WEC.

In summary, the null energy condition is always violated, both inside and outside the black hole. The strong energy condition is satisfied outside the black hole; however, inside it, it may be satisfied up to a certain value and then violated, or satisfied for all values depending on the parameter $a$. Finally, the weak energy condition is satisfied up to a certain value and then becomes violated, whereas inside the black hole it would tend to exhibit the same behavior, but assuming $\alpha<<1$, it is always satisfied.

\subsection{Field Sources for Regular Schwarzschild de Sitter Solution}\label{FSfRSsSS}

In the previous section, we analyzed the energy conditions by determining the energy density and the anisotropic pressures of the fluid that acts as the source of this solution. However, we did not address the nature of this source, that is, which fields can be used as a source for the metric defined in (\ref{5}).

The main candidates for sources of regular black hole or wormhole solutions are a phantom scalar field \cite{Bronnikov:2005gm} and nonlinear electrodynamics \cite{Ayon-Beato:1998hmi}, with a preference for magnetic sources. For static and regular solutions, it is possible to have either an electric or a magnetic source, but it is not possible to have a dyonic source, with both simultaneously. The electric source is more problematic because it does not recover the results of Maxwell electrodynamics in the weak-field limit, which would be expected for consistency in a nonlinear electrodynamics theory.

For black bounces, one has to consider both types of sources, a scalar field and a non-linear electrodynamics, simultaneously, as pointed in previous literature \cite{Bronnikov:2021uta, Alencar:2025jvl}. Hence, the following gravitational action is considered
\begin{eqnarray}
\nonumber S=&&\frac{1}{2\kappa^2}\int d^4x\sqrt{-g}\,(R-2\Lambda)\\ \nonumber &&-\int d^4x\sqrt{-g}\,(\epsilon g^{\alpha\beta}\partial_{\alpha}\Phi\partial_{\beta}\Phi+V(\Phi))\\ &&-\int d^4x\sqrt{-g}\,L(F)\ , \label{31}
\end{eqnarray}
where $\epsilon=\pm 1$, with the positive sign corresponding to a standard scalar field and the negative sign to a phantom scalar field; $\Phi$ represents the scalar field, $V(\Phi)$ its associated potential, and $L(F)$ is the Lagrangian of nonlinear electrodynamics, while $F=\frac{1}{4}F^{\mu\nu}F_{\mu\nu}$ is the Maxwell invariant (to recover Maxwell electrodynamics, one simply sets $L=F$), where $F_{\mu\nu}=\partial_\mu A_\nu-\partial_\nu A_\mu$.

By defining $T^{\mu}{}_{\nu}[\Phi]$ as the components of the stress–energy tensor associated with the scalar field, and $T^{\mu}{}_{\nu}[NED]$ as the components associated with nonlinear electrodynamics, the following equations of motion are obtained
\begin{eqnarray}
&&G^{\mu}{}_{\nu}+\delta^{\mu}_{\nu}\Lambda=\kappa^2(T^{\mu}{}_{\nu}[\Phi]+T^{\mu}{}_{\nu}[NED]);\label{32}\\ &&2\epsilon\nabla_{\mu}\nabla^{\mu}\Phi=\frac{dV(\Phi)}{d\Phi};\label{33}\\ &&\nabla_{\mu}(L_FF^{\mu\nu})=\partial_{\mu}(\sqrt{-g}L_FF^{\mu\nu})=0\ ,\label{34}
\end{eqnarray}
where $L_F\equiv\frac{\partial L}{\partial F}$. Let us assume a magnetic source, such that $F_{\theta\phi}=q_m\sin\theta$. In this case, the Maxwell invariant is given by
\begin{equation}\label{35}
F=\frac{q_m^2}{2(r^2+a^2)^2}\ .
\end{equation}

The scalar field is assumed to depend solely on the radial coordinate $r$. Then, the components of the stress–energy tensor for each source yield
\begin{eqnarray}
&&T^{\mu}{}_{\nu}[NED]=\frac{q_m^2L_F}{(r^2+a^2)^2}(0,0,1,1)-L(F)\delta^{\mu}_{\nu}\ ;\label{36}\\ &&T^\mu{}_\nu[\Phi]=\epsilon A(r)\Phi'(r)^2(-1,1,-1,-1)-V(r)\delta^\mu_\nu\ .\label{37}
\end{eqnarray}

By substituting the expressions (\ref{36}) and (\ref{37}) into the field equations (\ref{32}), the equations of motion lead to
\begin{eqnarray}
\nonumber &&a^2\frac{\left[\alpha^2(r^2+a^2)^{3/2}+\sqrt{r^2+a^2}-4M\right]}{(r^2+a^2)^{5/2}}=\\ &&-\kappa^2\Big(\epsilon A(r)\Phi'(r)^2+V(r)+L(r)\Big)\ ,\label{38}\\ \nonumber &&a^2\frac{[3\alpha^2(r^2+a^2)-1]}{(r^2+a^2)^2}=\\ &&\kappa^2\Big(\epsilon A(r)\Phi'(r)^2-V(r)-L(r)\Big)\ ,\label{39}\\ \nonumber && a^2\frac{\left[\alpha^2(r^2+a^2)^{3/2}+\sqrt{r^2+a^2}-M\right]}{(r^2+a^2)^{5/2}}=\\ &&-\kappa^2\left[\epsilon A(r)\Phi'(r)^2+V(r)+L(r)-\frac{q_m^2L_F(r)}{(r^2+a^2)^2}\right]\ .\label{40}
\end{eqnarray}

Subtracting equations (\ref{39}) and (\ref{38}), the scalar field can be isolated in the following equation
\begin{equation}\label{41}
\Phi'(r)^2=-\frac{a^2}{\epsilon\kappa^2(r^2+a^2)^2}\ .
\end{equation}

The lhs of (\ref{41}) must necessarily be positive, since the scalar field must be real. Then, the only way for the rhs to be also positive is such that $\epsilon=-1$, which cancels the overall negative sign in this equation. Hence, a phantom scalar field is required, which can be obtained by integrating (\ref{41}),
\begin{equation}\label{42}
\Phi(r)=\frac{1}{\kappa}\text{tg}^{-1}\left(\frac{r}{a}\right)\ .
\end{equation}

Here the boundary condition $\Phi(r=0)=0$ is assumed. 

The scalar potential $V(r)$ is obtained through (\ref{33}),
\begin{align}V'(r)={}&-2\bigg(A'(r)\Phi'(r)+A(r)\Phi''(r) \nonumber\\&\quad+\frac{2rA(r)}{r^2+a^2}\Phi'(r)\bigg)\Phi'(r),\label{43}\\V(r)={}&\frac{4a^2}{\kappa^2}\left(\frac{M}{5(r^2+a^2)^{5/2}}-\frac{\alpha^2}{2(r^2+a^2)}
\right).\label{44}
\end{align}

\begin{figure}[t!]
\centering
\includegraphics[scale=0.7]{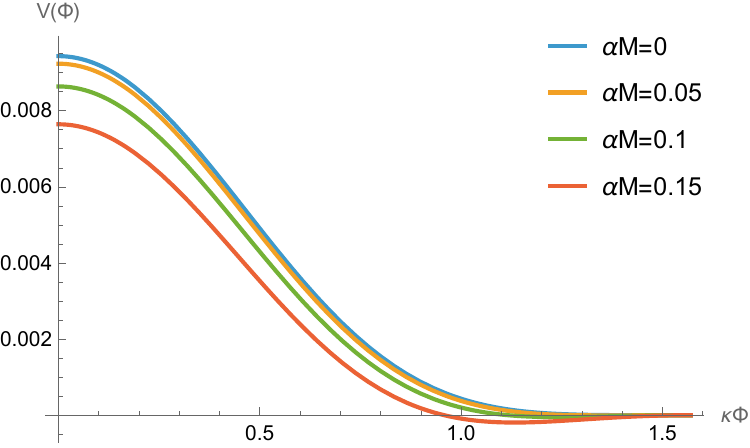}
\caption{The scalar potential as a function of the scalar field is shown for some values of $\alpha$.}
\label{Fig10}
\end{figure}

The potential can be expressed as a function of the field by isolating $r$ in terms of $\Phi$ and substituting this result into (\ref{44}), which leads to
\begin{equation}\label{45}
V(\Phi)=\frac{4M\cos^5(\kappa\Phi)}{5a^3\kappa^2}-\frac{2\alpha^2\cos^2(\kappa\Phi)}{\kappa^2}\ .
\end{equation}
These results show that the potential depends on trigonometric functions, specifically on powers of the cosine function. In this case, the cosmological constant influences the value of the potential with a dependence of $\cos^2(\kappa\Phi)$. In Fig.~\ref{Fig10}, the scalar potential $V(\Phi)$ is depicted.

Then, by equation (\ref{39}), one can calculate $L(r)$,
\begin{equation}\label{46}
L(r)=\frac{6Ma^2}{5\kappa^2(r^2+a^2)^{5/2}}\ ,
\end{equation}
which can be easily expressed in terms of $F$ by using (\ref{35}),
\begin{equation}\label{47}
L(F)=\frac{6Ma^2}{5\kappa^2q_m^{5/2}}(2F)^{5/4}\ .
\end{equation}

This result reveals an interesting behavior: it is not affected by the cosmological constant, that is, it is exactly the same as the one obtained for the Simpson–Visser regularization of asymptotically flat Schwarzschild spacetime \cite{Rodrigues:2023vtm}, which can be verified by noting that $L$ does not depend on $\alpha$. As pointed above, in the case of the scalar field potential, the cosmological constant does affect the behavior, adding an extra term to the expression for the potential, although it does not explicitly arise in the expression for the scalar field itself.

Finally, $L_F$ can be directly obtained through (\ref{40}),
\begin{equation}\label{48}
L_F(r)=\frac{3Ma^2}{\kappa^2q_m^2\sqrt{r^2+a^2}}\ .
\end{equation}
A consistency condition that must be satisfied is that $L_F(r)-L'(r)(F'(r))^{-1}=0$, since $L_F=\frac{\partial L}{\partial F}=\frac{dL}{dr}\frac{dr}{dF}$. By considering the results of (\ref{36}), (\ref{46}) and (\ref{48}), the consistency condition is satisfied.


\section{Geodesics and Optical Appearance of the Schwarzschild de Sitter Black Bounces} \label{GaOotSdSBB}
In this section, the optical appearance of the central object described by the metric (\ref{5}) is simulated. To do so, several steps are required, which will be developed throughout this section. Firstly, null geodesics followed by photons emitted by the accretion disk are obtained (a procedure known as ray tracing), where the effective potential of trajectories is analyzed to obtain the possible circular orbits as well as the scattering and capture points of photons. Then, the transfer functions are achieved, which relate the intersection of the geodesics with the accretion disk and the impact parameter of the trajectory. Finally, by assuming an intensity profile for the emission of the accretion disk, the luminosity as observed by a distant observer is calculated and the optical appearance (the shadow) is obtained. 

\subsection{Null Geodesics for Regular Schwarzschild dS Spacetime}
To determine the trajectories followed by photons emitted from the accretion disk of the central object, it is necessary to study the geodesics obtained from the metric defined in (\ref{5}). To this end, we parametrize the metric with the affine parameter $\tau$, which defines a given trajectory, and we also assume that the trajectories lie in the equatorial plane (i.e., $\theta=\pi/2$),
\begin{equation}\label{51}
\left(\frac{ds}{d\tau}\right)^2\equiv \epsilon=-A(r)\dot{t}^2+\frac{\dot{r}^2}{A(r)}+(r^2+a^2)\dot{\phi}^2\ ,
\end{equation}
where the dot denotes derivatives with respect to $\tau$. Here, $\epsilon=-1$ for timelike trajectories (massive particles) and $\epsilon=0$ for null trajectories (massless particles). By considering the Killing vectors of this spacetime, we obtain the following constants of motion
\begin{equation}\label{52}
E=A(r)\dot{t}\ ,\quad L=(r^2+a^2)\dot{\phi}\ .
\end{equation}

\begin{figure}[t!]
\centering
\includegraphics[scale=0.7]{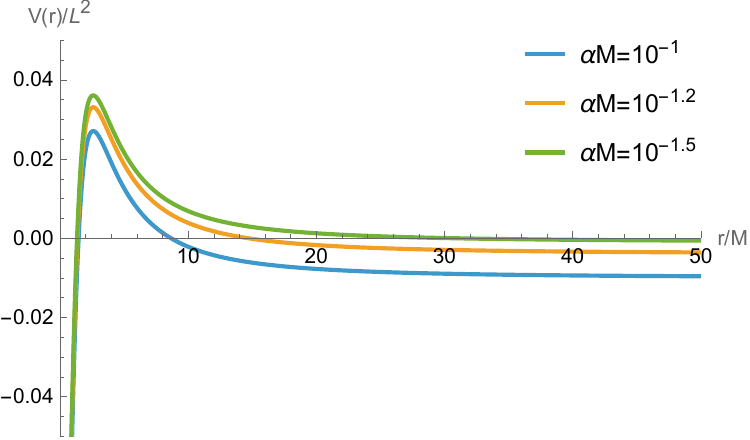}
\caption{Effective potential for null geodesics for the regular Schwarzschild-de Sitter metric. Here, we are considering different values for $\alpha$.}
\label{Fig13}
\end{figure}

It is important to note that, since our spacetime is not asymptotically flat, we cannot interpret $E$ as energy per unit mass and $L$ as angular momentum per unit mass in the case of massive particles.
Instead, $E$ and $L$ are simply parameters with dimensions of energy and angular  momentum (per unit mass), respectively.
By expressing $\dot{t}$ and $\dot{\phi}$ in terms of the constants of motion and substituting into (\ref{51}), the equation for null geodesics become
\begin{eqnarray}
&&\dot{r}^2=E^2-A(r)\left(\frac{L^2}{r^2+a^2}-\epsilon\right)\ ,\label{53}\\
&&V_{eff}(r)\equiv A(r)\left(\frac{L^2}{r^2+a^2}-\epsilon\right)\ , \label{54}
\end{eqnarray}
where $V_{eff}(r)$ is the effective potential of the trajectories (see Fig.~\ref{Fig13}), 
\begin{equation}\label{55}
V_{eff}(r)=\frac{L^2}{r^2+a^2}\left(1-\frac{2M}{\sqrt{r^2+a^2}}\right)-\alpha^2L^2.
\end{equation}

The term on the rhs of (\ref{55}) is precisely the effective potential of the usual Schwarzschild-like black bounce solution, with the addition of a constant term that depends on the cosmological constant. To obtain the trajectory equation, the impact parameter $b\equiv \frac{L}{E}$, is defined. Then, by using \eqref{52} and \eqref{53}, the equation for $\phi=\phi(r)$ yields
\begin{eqnarray}
    &&\frac{d\phi}{dr}=\pm \frac{b}{r^2+a^2}\left[1-\frac{b^2A(r)}{r^2+a^2}\right]^{-1/2}\ ,\label{57}
\end{eqnarray}
Now, let's evaluate the effective potential and determine some basic properties about the geodesics. Firstly, we are interested in getting the turning/scattering points $r_0$ i.e. those locations of the radial coordinate where $\dot{r}=0$. From (\ref{53}) and using the definition of impact parameter, the value of the impact parameter and $r_0$ are related, leading to   
\begin{equation}\label{58}
    b_0=\sqrt{\frac{r_0^2+a^2}{A(r_0)}}\ .
\end{equation}  
In addition, as far as $\ddot{r}= 0 =\dot{r}$, the photon is in a circular orbit, which is associated with the critical impact parameter
\begin{equation}\label{59}
    b_c=\sqrt{\frac{r_c^2+a^2}{A(r_c)}}\ ,
\end{equation}
whereas the location of the circular orbit $r_c$ is obtained through the condition $V'_{eff}=0$, which yields
\begin{equation}
    V'_{eff}=0\rightarrow \sqrt{r_c^2+a^2}=3M\ ,
\end{equation}
This result is the same as the case for the assymptotically flat black bounce, where $\alpha=0$. To know whether this circular orbit is unstable or stable, we need to evaluate the sign of the second derivative of the effective potential
\begin{equation}
    V''_{eff}(r_c)=-\frac{2L^2}{(3M)^6}r_c^2<0\ .
\end{equation}
Once the sign is negative, this circular orbit is unstable. By substituting the value of the circular orbit position in the expression (\ref{58}), the critical impact parameter turns out
\begin{equation}\label{62}
    b_c=\frac{3\sqrt{3}M}{\sqrt{1-27\alpha^2M^2}}\ ,
\end{equation}
which is larger than the critical value for the flat case, that's $3\sqrt{3}M$. 

More precisely, this can be easily seen by treating $  \alpha  $ as a small parameter and expanding around $\alpha=0$. In this way, the critical impact parameter $  b_c  $ can be approximately expanded as
\begin{align}
    b_c \approx 3\sqrt{3}M 
    \Bigg(
    1 + 
    \frac{27}{2}  \alpha ^2 M^2 +
    \frac{2187}{8} \alpha ^4 M^4 
    \Bigg).
\end{align}
For an impact parameter below the critical one, photons trajectories will end in the central object.

For asymptotically de Sitter spacetimes, the effective potential might become negative, as shown Figure \ref{Fig13}. In this case, $E$ might be any positive value. However, the minimum value for $E$ has to be the same as $V_{eff}$ at the observer position, and the observer may be at a point where $V_{eff}>0$. This is true because the observer location needs to be closer than the cosmological horizon, where the effective potential is necessarily positive. This condition is not necessary in asymptotically flat spacetimes because there is no cosmological horizon and one can assume that the observer is located at infinity, where $V_{eff}\rightarrow 0$.

As a consequence, a maximum value for the impact parameter arises. By approximating $A(r)\approx 1-\alpha^2r^2$ for large values of $r$, the maximum value for the impact parameter yields
\begin{equation}\label{63}
    b_{max}\approx \frac{r_{obs}}{\sqrt{1-\alpha^2r_{obs}^2}}\ .
\end{equation}
Since $r_{obs}$ might be very large, the maximum value of the impact parameter becomes large too. 

\subsection{Transfer Function}
The optical appearance of a compact object depends on several factors that involve the physics of the accretion disk and the geometry of the background spacetime. However, the physics of the accretion disk is not well-known and very complex, including absorptivity, scattering, and emission, among other effects that are known to play a heavy role, such as the behaviour of magnetic fields and turbulence, requiring the development of sophisticated radiative transfer codes \cite{EventHorizonTelescope:2020tst}. For this work, let's consider a simplified model of the emission of photons, where we won't need to do a rigorous and complex analysis. But, to begin this analysis, it is necessary to study the so-called transfer functions.

To generate the images, it is first necessary to determine the emission points on the accretion disk as functions of the impact parameter for the different emission orders, namely, the number of turns around the central object a photon might undergo. Following previous conventions, we will call direct emission ($n=0$) for those photons emitted at some radial distance from the object that go directly to the observer, the lensed emission ($n=1$) for those photons that cross the accretion disk once after emitted and the photon ring emission for those crossing the accretion disk more than once ($n\geq 2$). This information defines the transfer function, which consists on the curves corresponding to these emission modes that relate the location where the photon is emitted with the impact parameter for each type of emission. Moreover, the slopes of these curves are directly related to the degree of demagnification of the corresponding images \cite{Perlick:2021aok}, which are highly suppressed for increasing orders on $n$, such that here only orders up to $n=2$ is considered. In the canonical black hole case, the $n=1$ and $n=2$  curves are concentrated close to the critical impact parameter $b_c$, reflecting their close connection with the photon sphere. 

\begin{figure*}[t!]
\centering
\includegraphics[scale=0.5]{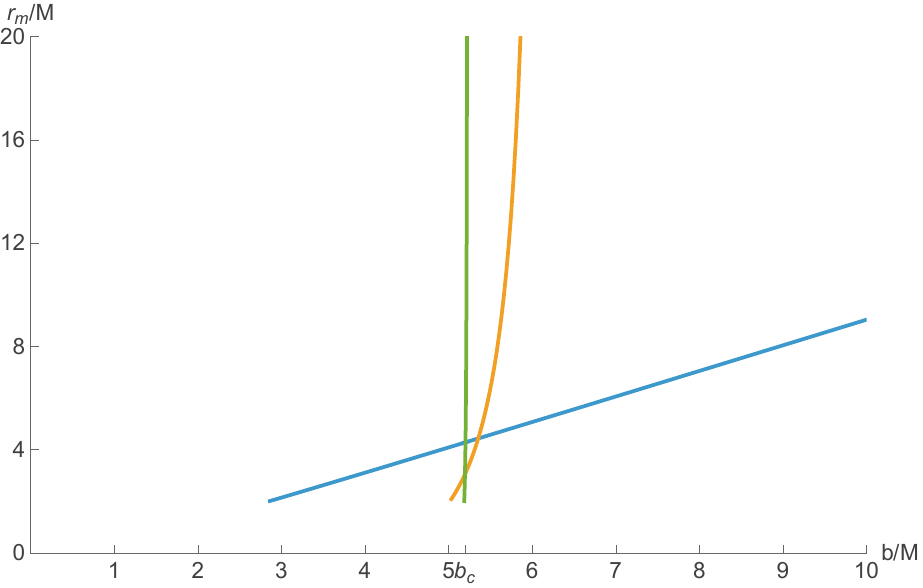}
\includegraphics[scale=0.5]{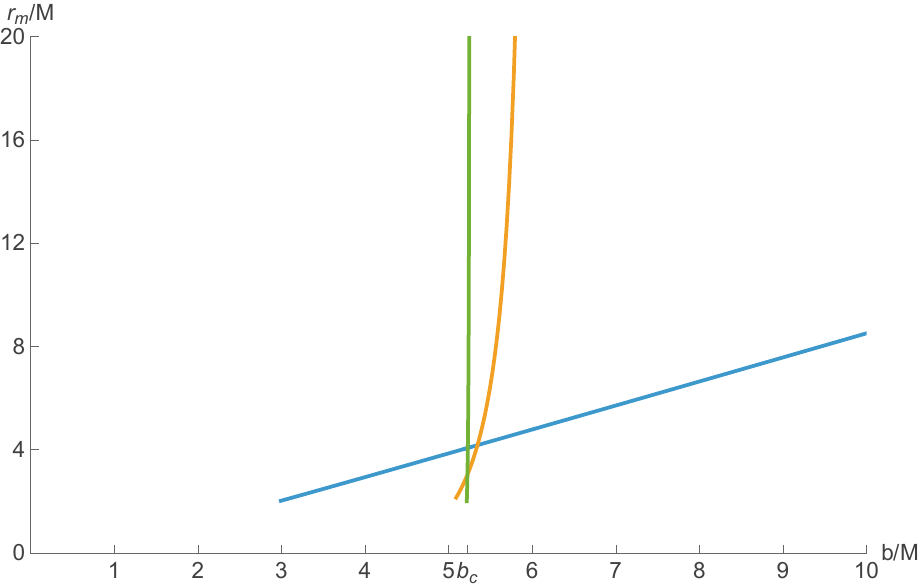}
\includegraphics[scale=0.45]{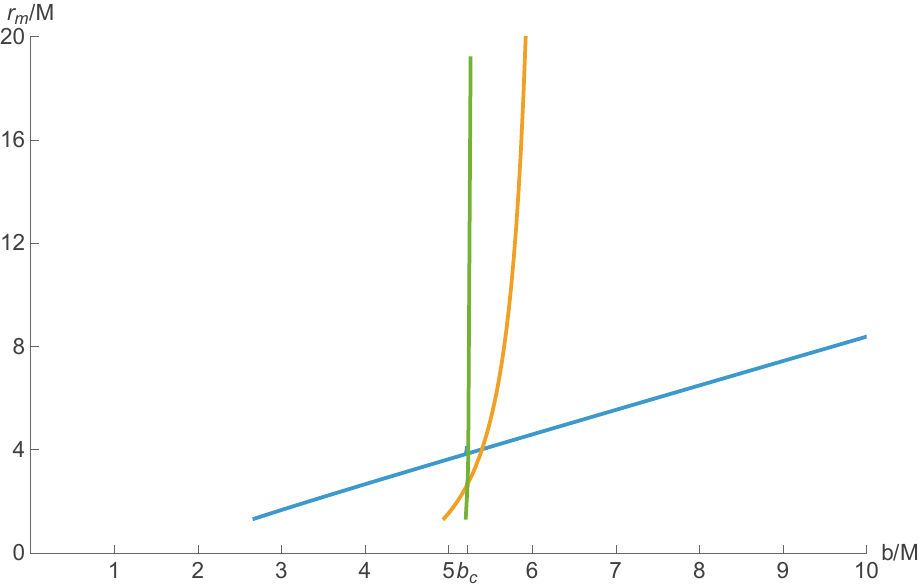}
\includegraphics[scale=0.4]{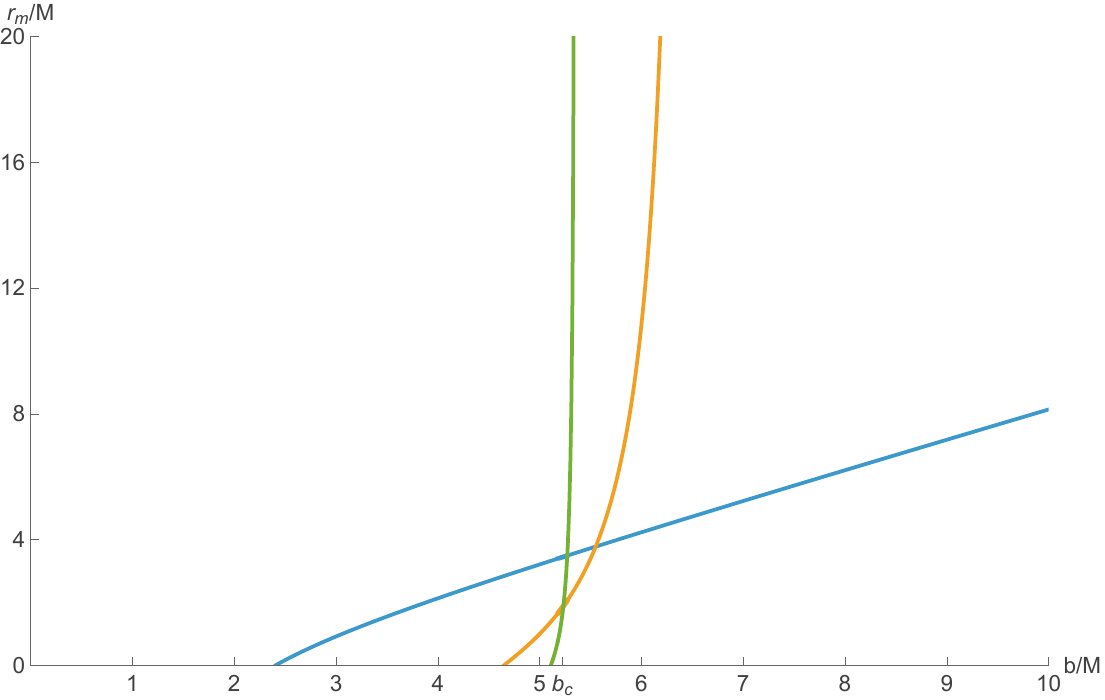}
\caption{Transfer functions. For every panel: the direct emission ($n=0$) is depicted by the blue curve, the lensed emission ($n=1$) is given by orange and the photon ring ($n=2$) by green. Top panels: flat Schwarzschild spacetime (left), Schwarzschild-de Sitter metric (right). Bottom panels: regular black hole with $a=1.5$ (left) and transversable wormhole with $a=2.5$ (right). Here, the cosmological horizon is located at $x_{CH}=\sqrt{r_c^2+a^2}=40M$. }
\label{Transfer1}
\end{figure*}

\begin{figure*}[t!]
\centering
\includegraphics[scale=0.3]{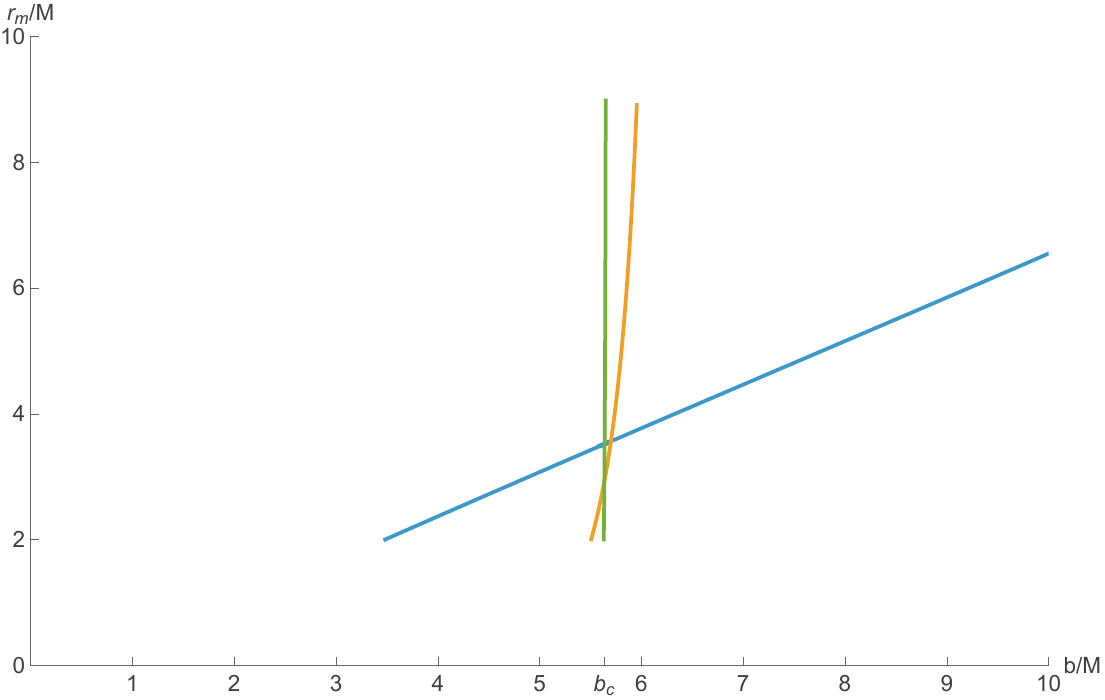}
\includegraphics[scale=0.3]{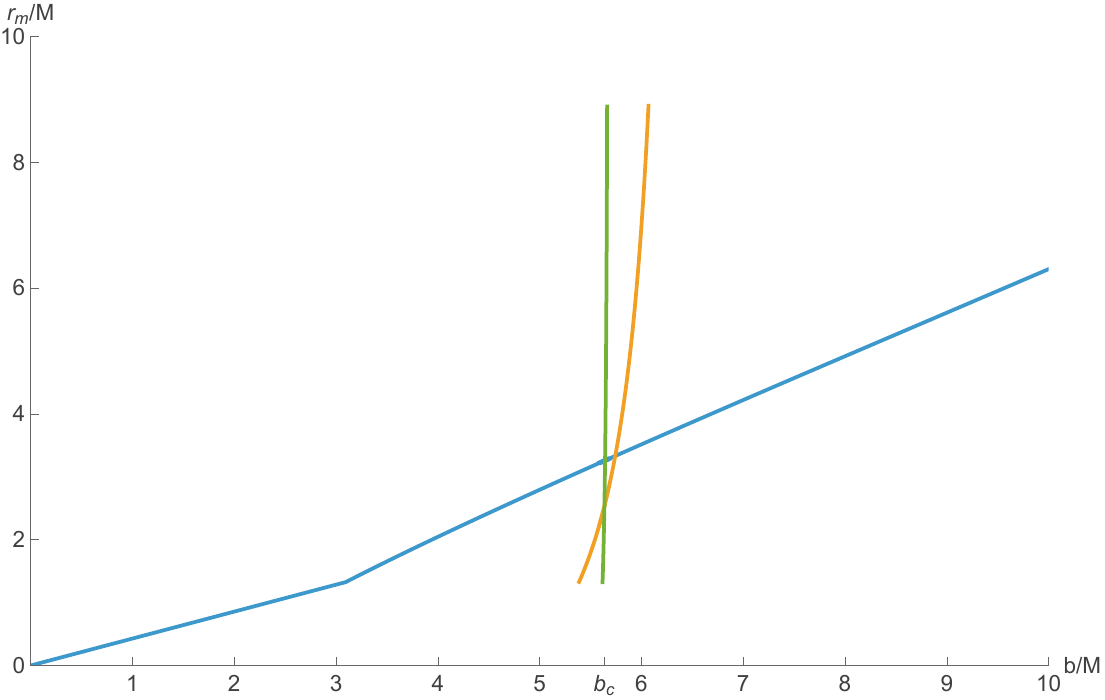}
\includegraphics[scale=0.3]{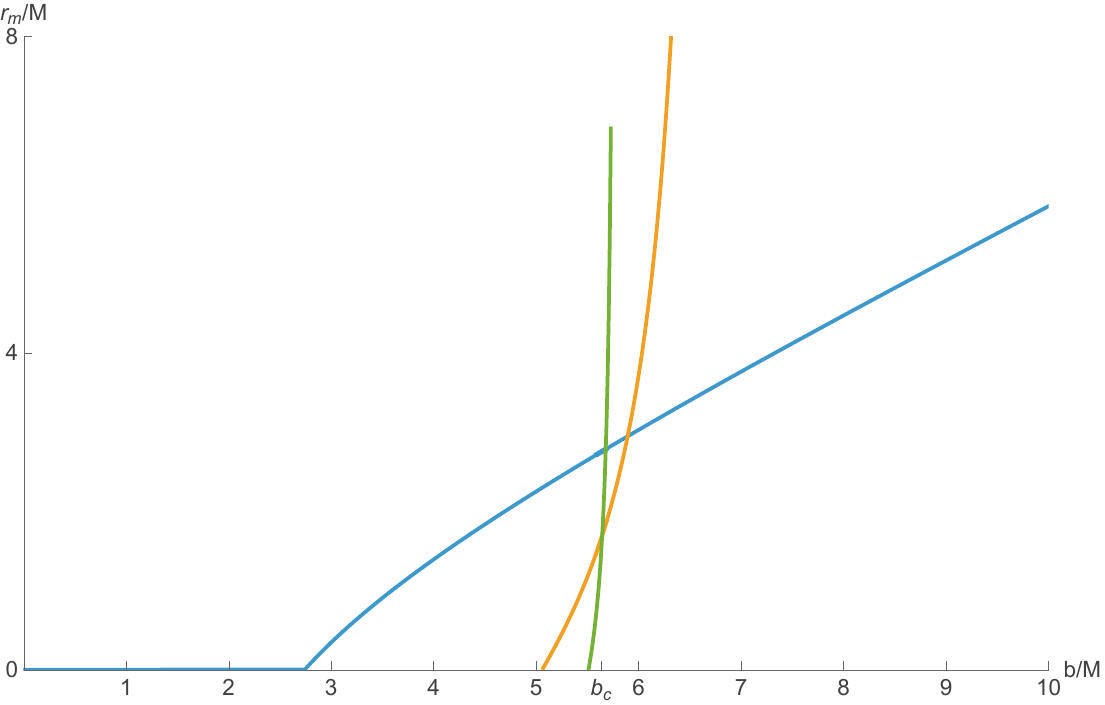}
\caption{Transfer functions for the the direct emission (blue), the lensed emission (orange) and the photon ring emission (green). From left to the right: Schwarzschild-de Sitter spacetime, a regular black hole with $a=1.5$ and a transversal wormhole with $a=2.5$. Here, the cosmological horizon is located at $x_{CH}=\sqrt{r_c^2+a^2}=10M$.}
\label{Transfer2}
\end{figure*}

In Figs.~\ref{Transfer1} and \ref{Transfer2}, the transfer functions are depicted for several cases. The usual assymptotically flat Schwarzschild spacetime is also included, as well as Schwarzschild-de Sitter metric, a regular de Sitter black hole and a de Sitter wormhole.

\subsection{Optical appearance} 

\begin{figure}[t!]
\centering
\includegraphics[scale=0.5]{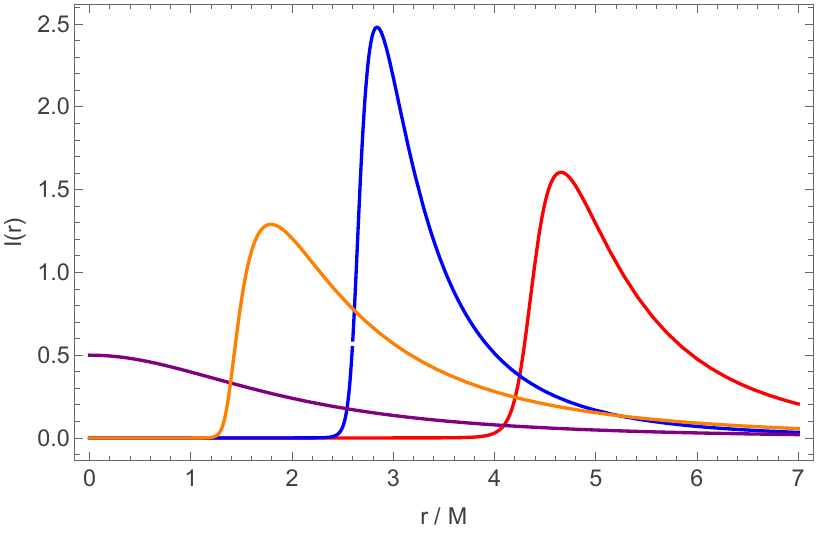}
\caption{Some samples of the intensity profile (\ref{IntensitySU}) for different values of the parameters $\{\mu, \sigma, \gamma\}$.}
\label{intensity_samples}
\end{figure}

\begin{figure*}[t!]
\centering
\includegraphics[scale=0.52]{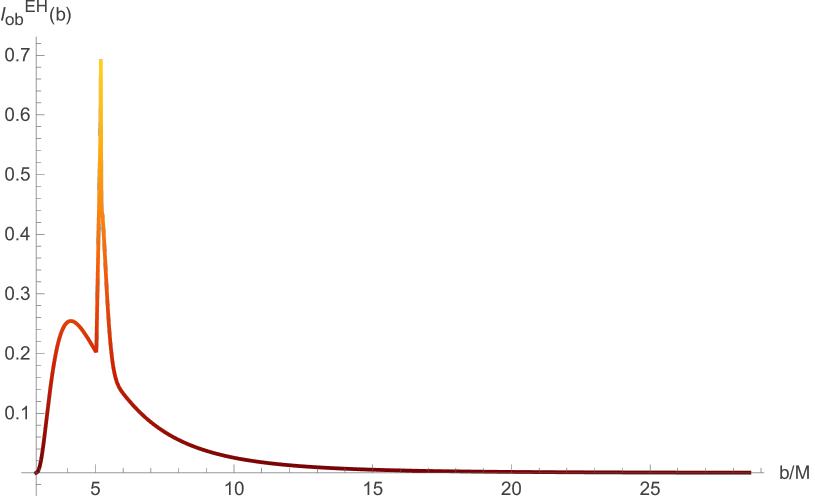}
\includegraphics[scale=0.5]{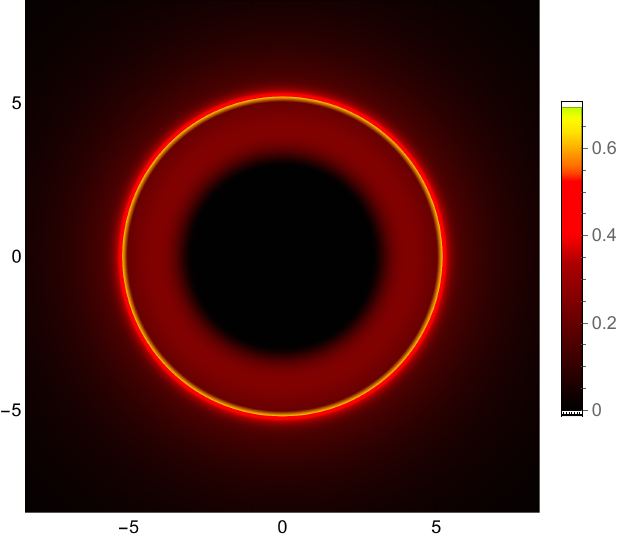}
\includegraphics[scale=0.47]{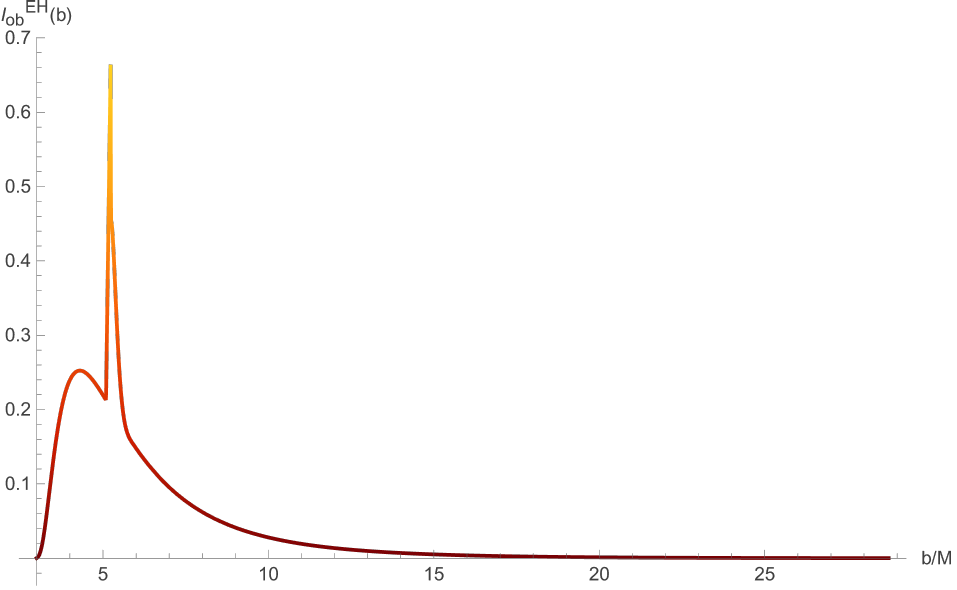}
\includegraphics[scale=0.5]{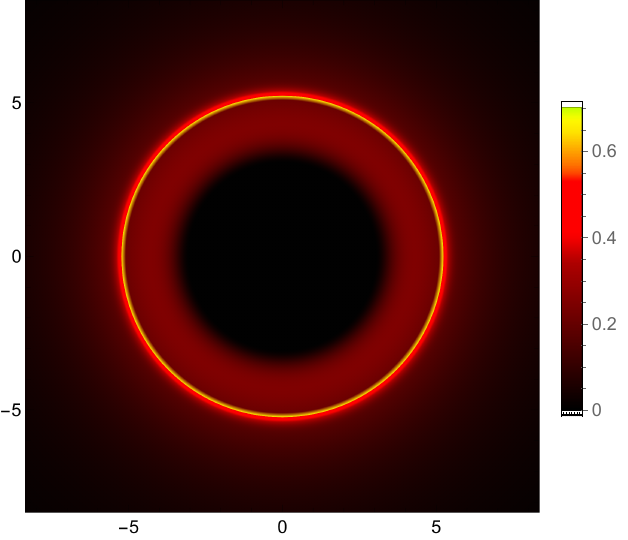}
\includegraphics[scale=0.5]{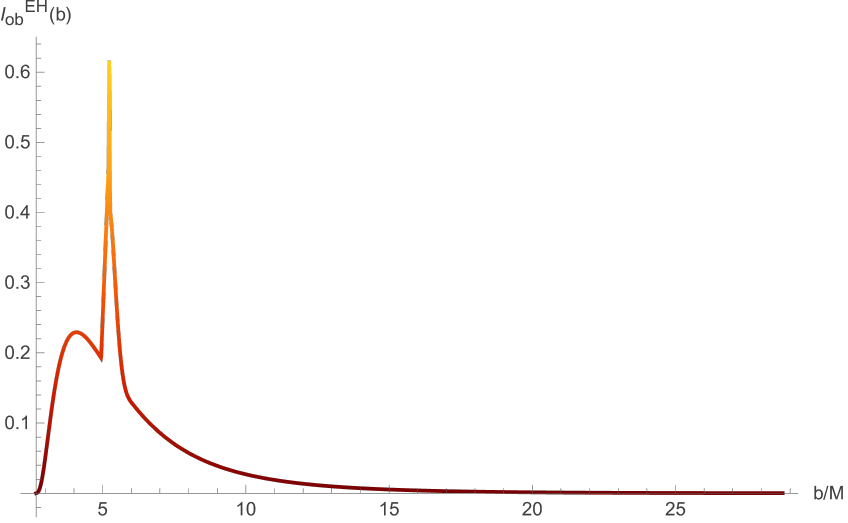}
\includegraphics[scale=0.5]{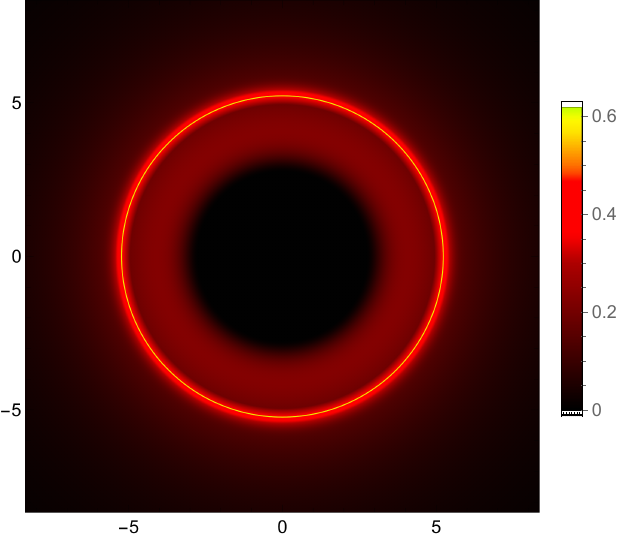}
\includegraphics[scale=0.5]{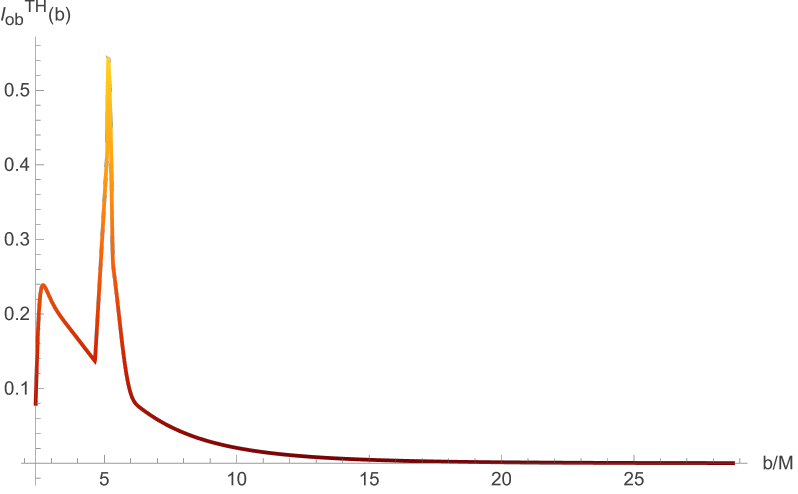}
\includegraphics[scale=0.5]{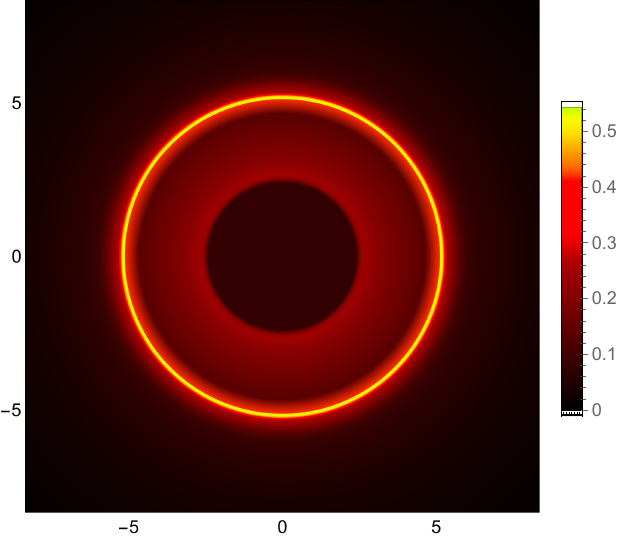}
\caption{Observed intensities when considering the intensity profile (\ref{IntensitySU}) with $\gamma=-2$, $\mu=r_{EH/TH}$ and $\sigma=1/4$. From top to the bottom: the usual Schwarzschild black hole, the Schwarzschild-de Sitter spacetime, a regular de Sitter black hole with $a=1.5$ and a de Sitter wormhole with $a=2.5$. The cosmological horizon is located at $x_{CH}=\sqrt{r_c^2+a^2}=40M$ for every case where it applies, while the event horizon, if any, is located at $x_{EH}=\sqrt{r_c^2+a^2}=2M$.}
\label{Shadows1}
\end{figure*}

\begin{figure*}[t!]
\centering
\includegraphics[scale=0.5]{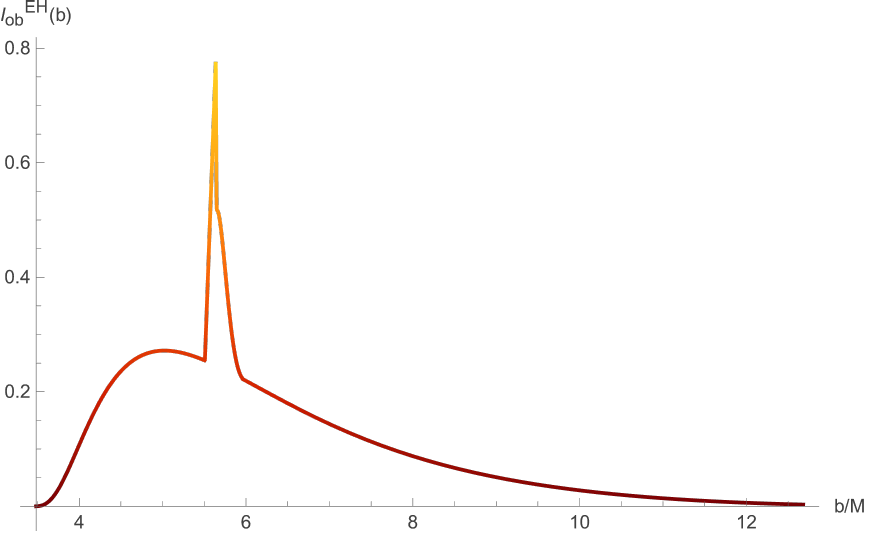}
\includegraphics[scale=0.5]{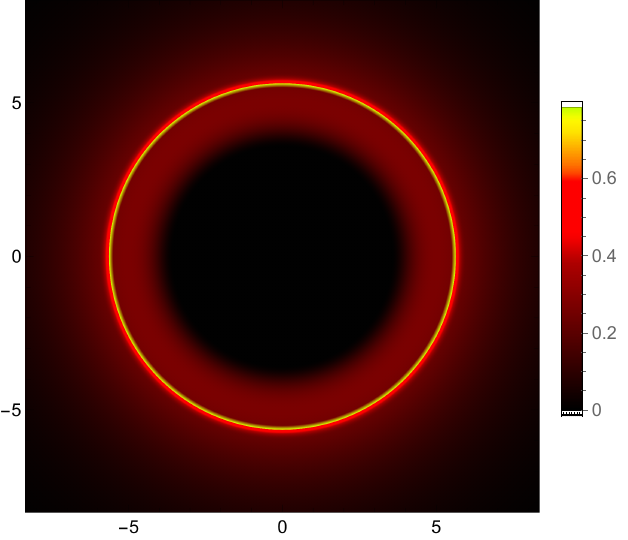}
\includegraphics[scale=0.6]{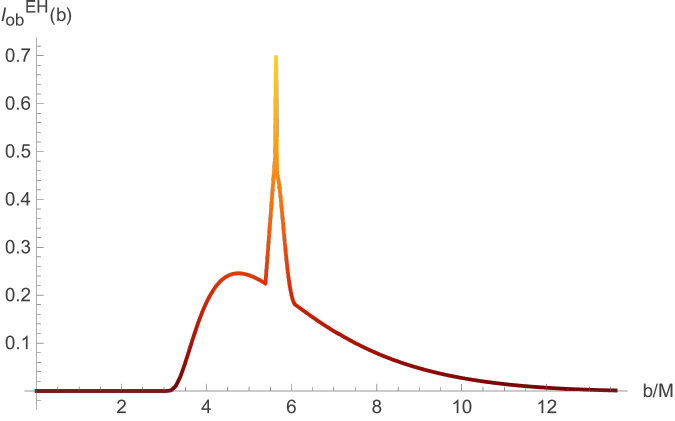}
\includegraphics[scale=0.5]{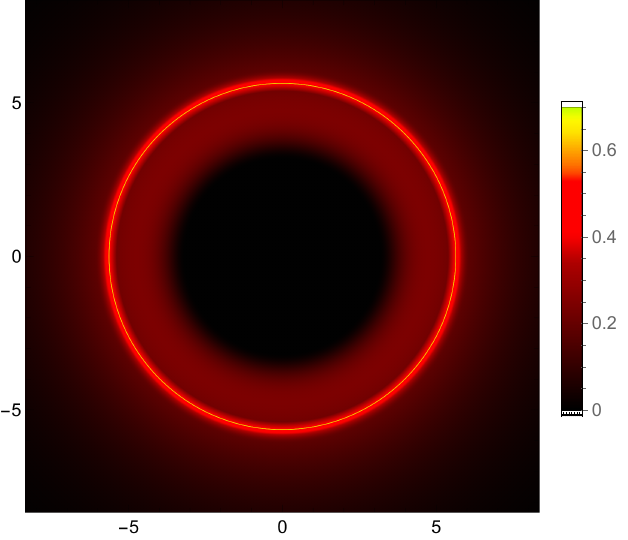}
\includegraphics[scale=0.5]{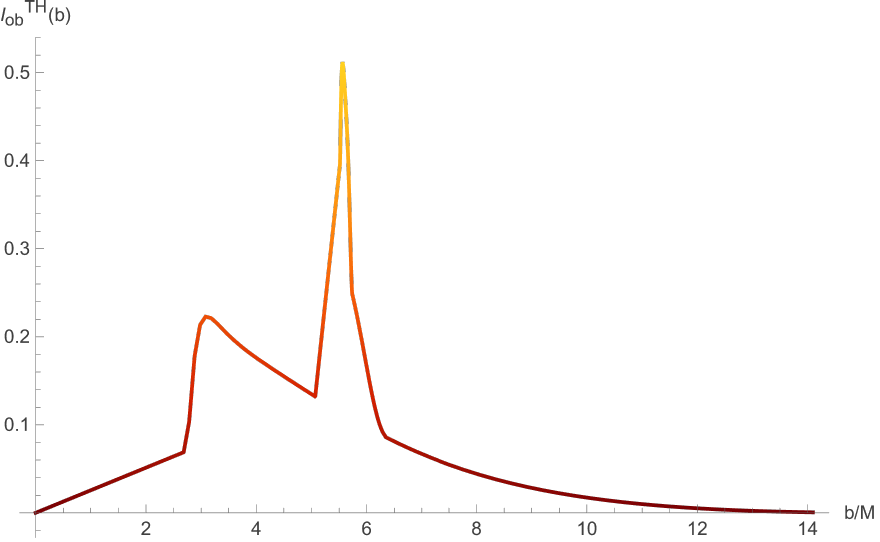}
\includegraphics[scale=0.5]{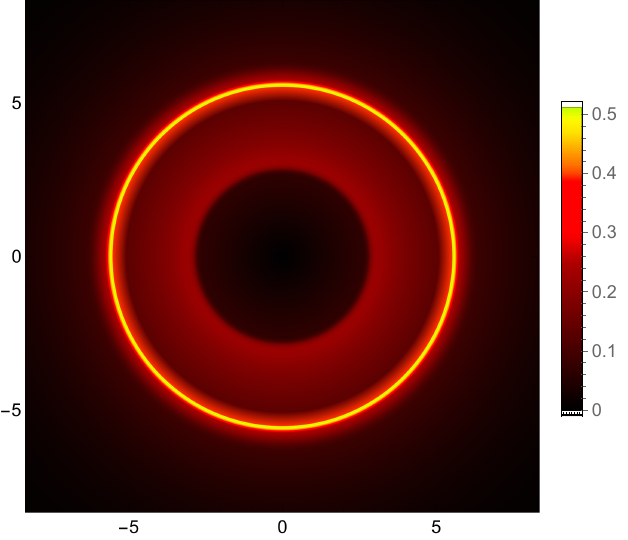}
\caption{Observed intensities for wheh considering the intensity profile (\ref{IntensitySU}) with $\gamma=-2$, $\mu=r_{EH/TH}$ and $\sigma=1/4$. From top to the bottom: flat Schwarzschild black hole, Schwarzschild-de Sitter spacetime, regular de Sitter black hole with $a=1.5$ and de Sitter wormhole with $a=2.5$. The cosmological horizon is located at $x_{CH}=\sqrt{r_c^2+a^2}=10M$ for every case where it applies, and the event horizon located at $x_{EH}=\sqrt{r_c^2+a^2}=2M$, if any.}
\label{Shadows2}
\end{figure*}

To the end of simulating the optical appearance for the objects described by a regular Schwarzschild-de Sitter spacetime, a particular luminosity profile for the emission of the accretion disk is assumed. Firstly, the approximations of an optically and geometrically thin disk are considered. Then, the accretion disk is located perpendicular to the line of sight while the intensity is assumed to be isotropic in frequency $\nu$, so that the emitted intensity depends only on the radial coordinate $I^{em}_{\nu} = I_{\nu}(r)$, neglecting absorption effects, making $I_{\nu}/\nu^3$ invariant along the photon path \cite{Mihalas:1984}. As photons frequency moving in a non-flat spacetime is affected by redshift, the intensity at an observed point leads to,
\begin{equation}
I^{ob}_{\nu'} = \left(\frac{A(r)}{A(r_{ob})}\right)^{3/2} I^{em}_{\nu}\ ,
\end{equation} 
while the total intensity is obtained by integrating over all frequencies,
\begin{equation}
I^{ob} = \int I^{ob}_{\nu'}\, d\nu' = \left(\frac{A(r)}{A(r_{ob})}\right)^2 I(r)\ .
\end{equation}
However, each intersection of photons trajectories with the accretion disk peaks some addition intensity for a given impact parameter such that the intensity at the observer screen is given by
\begin{equation}
    \left. I^{obs}(b)=\frac{1}{A(r_{ob})^2}\sum_n A^2(r)I(r)\right|_{r=r_n(b)} \ ,
\end{equation}
where $r_n(b)$ are the transfer functions for the $n-$th intersection with the equatorial plane. The final step for simulating the optical appearance consists on assuming a particular emission profile. In this sense, a widely used in previous literature, the so-called Standard Unbound (SU) distribution is considered, 
\begin{equation}
    I^{em}_{SU}(r,\mu,\sigma,\gamma)=\frac{e^{-\frac{1}{2}\left[\gamma+\text{arc
    sinh}\left(\frac{r-\mu}{\sigma}\right)\right]^2}}{\sqrt{(r-\mu)^2+\sigma^2}}\ .
        \label{IntensitySU}
\end{equation}
Here, the parameters $\mu$, $\sigma$, and $\gamma$ are the position of the emission peak, the scale (width) and the size of the central region, respectively. A set of samples is depicted in Fig.~\ref{intensity_samples}. In order to try to homogenize the different cases to be compared among them, the same width and scale are considered, whereas the peak of emission is always located at the event horizon of the black hole or at the throat for the wormhole case. Figures \ref{Shadows1} and \ref{Shadows2} show the intensity for an observer that is located close to the cosmological horizon for different types of objects, from the usual Schwazrschild-de Sitter black hole, a regular black hole and a wormhole, all of them asymptotically de Sitter. The asymptotically flat Schwarzschild spacetime is also included in Fig.~\ref{Shadows1} for comparison. The difference between Figs.~\ref{Shadows1} and \ref{Shadows2} is the location of the cosmological horizon and consequently its size in comparison to the event horizon. 

As shown, there are tiny differences between the different types of black holes, even in comparison to the usual flat Schwarzschild metric when the cosmological horizon is located far away. This is just because the asymptotic flatness is recovered when the cosmological horizon tends to infinity. However, by comparing both figures, \ref{Shadows1} and \ref{Shadows2}, one notes that the closer is the cosmological horizon, the smoother the intensity turns out, as shown by the slower decay in intensity in Fig.~\ref{Shadows2} in comparison to the case where the cosmological horizon is set further Fig.~\ref{Shadows1}.

\section{Quasi-normal Modes} \label{QNMs}

For the sake of this section, let us define again $x=\sqrt{r^2+a^2}$. The line element can now be rewritten as

\begin{equation} \label{eq:lineel}
    ds^2=-A(x)dt^2+\frac{dx^2}{B(x)}+\Sigma^2(x)\left(d\theta^2+\sin \theta d\phi^2\right)\ ,
\end{equation}
where,
\begin{equation}
    A(x)=B(x)=1-\frac{2M}{x}-\alpha^2x^2, \quad \Sigma(x)=x\ .
\end{equation}
For the axial gravitational perturbations are characterized by the Regge-Wheeler equation \cite{Regge:1957td}, which is a Schrödinger-like equation (for the derivation, see \cite{Duran-Cabaces:2025sly}) given by
\begin{align}
   \left [\partial_{r*}^2-\left(\omega^2+V_l(x) \right)\right]\phi(x,\omega)=0\ ,
    \label{eq_quasinormal}
\end{align}
where,
\begin{eqnarray} \label{eq:RWpot}
    V_l(x)&=&A(x)\Bigg[ \frac{l(l+1)-2}{\Sigma(x)^2} \\
    &+&\Sigma(x)\sqrt{\frac{B(x)}{A(x)}}\partial_x \left(\sqrt{A(x)B(x)}\partial_x\left(\frac{1}{\Sigma(x)}\right)\right)\Bigg]\ . \nonumber
\end{eqnarray}

and $r*$ is the well-known tortoise coordinate, which maps the region $r \in (r_{EH},+\infty)$ into $r* \in (-\infty,+\infty)$.

Solutions to equation \eqref{eq_quasinormal}, together with the corresponding boundary conditions, give the quasinormal modes (QNMs) spectrum of the spacetime described in \eqref{eq:lineel}. 

Notice that QNMs are distinguished from the normal modes (NMs) of a conservative system by the boundary conditions imposed on the linearized perturbations. Normal modes, described by real eigenfrequencies and representing undamped oscillations, are quite different from QNMs, which arise as a consequence of radiative boundary conditions: purely ingoing waves at the event horizon and purely outgoing waves at spatial infinity. As a result, their frequencies are complex, $  \omega_n=\omega_{R,n}+i\omega_{I,n}  $ with $  \omega_{I,n}<0  $ for a linearly stable black hole.
Moreover, the corresponding time dependence, $  e^{-i\omega_{R,n}t}\,e^{-|\omega_{I,n}|t}  $, describes exponentially damped oscillations. QNMs therefore encode the characteristic dissipative response of a black hole to external perturbations and dominate the late-time ringdown signal, rather than representing the stationary oscillations of a closed conservative system. As it is well known, QNMs are non-stationary states. Instead, they decay in time due to radiation at infinity and at the horizons (if any). For this reason, every mode has an associated vibrational complex frequency $\omega=\omega_R +i \omega_I$, where $\omega_R$ determines the oscillation frequency while $\omega_I$ describes the damping rate. The fundamental mode is defined as the least damped (and consequently, the longest-lived) mode, while the overtones are ordered according to increasing values of $|\omega_I|$.

The boundary conditions for this asymptotically de Sitter class of spacetimes depend on the physical scenario of interest. While the cosmological horizon exists, the existence of the event horizon depends on the value of the parameter $a$. If an event horizon exists, purely ingoing QNMs at the horizon ($r^*\rightarrow -\infty$) and purely outgoing QNMs at the cosmological horizon ($r^*\rightarrow \infty$) are imposed, leading to 
\begin{equation}
    \phi(r^*\rightarrow-\infty,\omega)\sim e^{-i\omega r^*},\ \phi(r^*\rightarrow\infty,\omega)\sim e^{i\omega r^*} \ .
    \label{eq_boundary}
\end{equation}
Moreover, in the absence of an event horizon and as a consequence of the potential symmetry in the tortoise coordinate, there will be symmetric ($\phi(-r^*,\omega)=\phi(r^*,\omega)$) and antisymmetric ($\phi(-r^*,\omega)=-\phi(r^*,\omega)$) QNMs. These QNMs will be purely outgoing at the cosmological horizon ($r^*\rightarrow \infty$) regardless of their parity, whereas at the throat, they become
{\small
\begin{equation}
    \begin{cases}
\frac{d\phi(r^*,\omega)}{dr^*}|_{r^*=r^*_\text{throat}}=0 & \text{if } \phi(r^*,\omega) \text{ symmetric}, \\
\phi(r^*_\text{throat},\omega)=0 & \text{if } \phi(r^*,\omega) \text{ antisymmetric},
\end{cases}
\end{equation}}
Several methodologies to compute QNM either numerically or semi-analytically have been proposed (see reviews of methodologies in \cite{Konoplya:2011qq, Pani:2013pma,Franchini:2023eda}). Direct integration and time domain analysis are numerical procedures that work well in order to find QNMs. Direct integration can help to find the fundamental mode and the overtones if a convenient initial guess, while time domain analysis allows to observe dynamical behaviour such as echoes in addition to obtaining the fundamental mode (for more information on this methodology see \cite{Duran-Cabaces:2025sly}). In this paper, we will use the pseudospectral method, which is a semianalytical methodology that works well to find the fundamental mode and overtones if one does not have an initial guess for them. 

\subsection{Pseudospectral method}

The main idea is to solve Eq.~\eqref{eq_quasinormal} by transforming the differential eigenvalue problem into a matrix eigenvalue problem, which can then be handled using standard algebraic methods. We closely follow the procedure described in \cite{Jansen2017}. Since the Cardinal functions associated with Chebyshev polynomials are naturally defined on the interval $[-1,1]$, the domain must ultimately be compactified to this range. However, for convenience, we first map the physical domain $x\in[x_{\text{inf}},x_{CH}]$ to the unit interval $u\in[0,1]$ through the change of variables.
\begin{equation}
    u=\frac{\frac{1}{x}-\frac{1}{x_{CH}}}{\frac{1}{x_\text{inf}}-\frac{1}{x_{CH}}}\ .
    \label{udefinition}
\end{equation}
Note that $x_\text{inf}$ can be either $x_{EH}$ or $a$, depending on the spacetime we are working with (with event horizon or without it, respectively). Then, the equation reads
\begin{align}
   \Big[H(u)^2\partial_{u}^2+H'(u)H(u)\partial_{u} \nonumber\\
   +\left(\omega^2-V^{RW}_l(x(u))\right)\Big]\phi(x(u),\omega)=0\ ,
   \label{eq_Hdiff}
\end{align}
where $H(u)$ is defined as
\begin{equation}
    H(u)=\frac{du}{dr^*}=\frac{du}{dx}\frac{dx}{dr^*}\ .
\end{equation}
For the spacetime \eqref{eq:lineel} and by definition \eqref{udefinition}, it gives
\begin{align}
&H(u)=\frac{
u \,\big(u x_{EH} x_{CH} - x_{\text{inf}}\big((u-1)x_{EH} + x_{CH}\big)\big)
}{
x_{CH}\, x_{\text{inf}}^2 \,(x_{EH}^2 + x_{EH} x_{CH} + x_{CH}^2)
}
 \nonumber\\[6pt]
&\quad \cdot\ 
\big(u (x_{EH}+x_{CH})(x_{CH} - x_{\text{inf}}) + x_{\text{inf}}(x_{EH} + 2x_{CH})\big)
\nonumber\\[6pt]
&\quad \cdot\ 
\sqrt{
1 - \frac{s^2 x_{EH}^2 \big(u(x_{CH} - x_{\text{inf}})+x_{\text{inf}}\big)^2}
{x_{CH}^2 x_{\text{inf}}^2}
}\ ,
\end{align}
where we used that 

\begin{equation}
    \alpha^2=\frac{1}{x_{EH}^2+x_{EH}x_{CH}+x_{CH}^2},
\end{equation}
and
\begin{equation}
    M=\frac{x_{EH}x_{CH}(x_{EH}+x_{CH})}{2(x_{EH}^2+x_{EH}x_{CH}+x_{CH}^2)},
\end{equation}
being $x_{EH}=\sqrt{a^2+r_{EH}^2}$ and $x_{CH}=\sqrt{a^2+r_{CH}^2}$ the corresponding horizons in the $x$ coordinate, as we defined in Eq.\eqref{12} and Eq. \eqref{13} . The next step is to impose the appropriate boundary conditions. For this reason, the following ansatz that encodes the behavior of the wave function at the boundaries is assumed
\begin{equation}
    \phi(x(u),\omega)=\frac{u^c}{(1-u)^b}h(u)\ ,
    \label{eq_ansatzbc}
\end{equation}
where $h(u)$ is a regular function for the whole domain, $u\in[0,1]$. The exponents $b$ and $c$ depend on the specific boundary conditions of each type of spacetime. For the case with an event horizon, Eq.\eqref{eq_boundary} turns out
\begin{equation}
    b=\frac{i\omega}{H'(1)}, \quad c=\frac{i\omega}{H'(0)}\ ,
    \label{eq_bcbh}
\end{equation}
whereas for the event horizonless case it reads
\begin{equation}
    b=\begin{cases}
0 & \text{if } \phi(u) \text{ symmetric} \\
-\frac{1}{2} & \text{if } \phi(u) \text{ antisymmetric}
\end{cases}\ , \quad c=\frac{i\omega}{H'(0)}\ .
\end{equation}
The last step is to perform a change of variables to adapt the domain to the Cardinal Functions domain $z\in[-1,1]$. Then, by the definition $z=2u-1$, the following equation is obtained
\begin{equation}
    c_2(z,\omega)k''(z)+c_1(z,\omega)k'(z)+c_0(z,\omega)k(z)=0\ ,
    \label{eq_coefdif}
\end{equation}
where the functions $c_2$, $c_1$, and $c_0$ can be expanded as
\begin{align}
c_2(z,\omega)=c_{22}(z)\omega^2+c_{21}(z)\omega+c_{20}(z)\ , \nonumber\\
c_1(z,\omega)=c_{12}(z)\omega^2+c_{11}(z)\omega+c_{10}(z)\ ,\label{eq_funcionsc}\\
c_0(z,\omega)=c_{02}(z)\omega^2+c_{01}(z)\omega+c_{00}(z)\ , \nonumber
\end{align}
and the explicit form for them is written in the Appendix. At this point is where the actual pseudospectral method starts. A grid of $N+1$ points is defined as 
\begin{equation}
    z_j=\cos \left(\frac{\pi \cdot j}{N}\right) \quad \text{for} \quad j=0,\dots ,N\ .
\end{equation}
The function $k(z)$ is also expanded in a linear combination of Cardinal functions $C(j,z)$,
\begin{equation}
    C(j,z)=\frac{2}{N p_j}\sum_{m=0}^N \frac{1}{p_m}T(m,x_j)T(m,x)\ ,
\end{equation}
being $T(n,z)$ the $n^{th}$ Chebyshev Polynomial in the variable $z$, and $p_0=p_N=2$, while $p_k=1$ for $k\neq0,N$.
\begin{equation}
    k(z)=\sum_{n=0}^N a_n\ C(n,z)\ .
\end{equation}
By using the derivation properties of the Chebyshev Polynomials and by defining the matrices
\begin{align}
    \hat{M}_2=c_{22}(z)\hat{D}^2+c_{12}(z)\hat{D}+c_{02}(z) \nonumber\\
    \hat{M}_1=c_{21}(z)\hat{D}^2+c_{11}(z)\hat{D}+c_{01}(z) \\
    \hat{M}_0=c_{20}(z)\hat{D}^2+c_{10}(z)\hat{D}+c_{00}(z) \nonumber
\end{align}
equation \eqref{eq_coefdif} can be written as
\begin{equation}
\left(\hat{M_2}\omega^2+\hat{M_1}\omega+\hat{M_0}\right)\vec{a}=0\ .
\label{eq_matrixeqnl}
\end{equation}
Here, $\vec{a}=(a_0,\dots,a_N)^T$ and $\hat{D}$ is \cite{Jaramillo:2020tuu}
\begin{equation}
    \hat{D}_{ij}=\begin{cases}
-\frac{z_i}{2(1-z_i^2)} & \text{if } i=j\neq0,N\ , \\
\frac{2N^2+1}{6} & \text{if } i=j=0\ , \\
-\frac{2N^2+1}{6} & \text{if } i=j=N\ , \\
\frac{\mathbb{c}_i}{\mathbb{c}_j}\frac{(-1)^{i+j}}{z_i-z_j} & \text{if } i\neq j\ , 
\end{cases}
\end{equation}
where
\begin{equation}
    \mathbb{c}_i=\begin{cases}
        2 & \quad \text{if} \quad j=0,N\ ,\\
        1 & \quad \text{if} \quad j\neq0,N\ .
    \end{cases}
\end{equation}
Since Eq. \eqref{eq_matrixeqnl} is not a linear equation in $\omega$, we can transform the matrix system $(N+1)\times(N+1)$ into an equivalent matrix system $2(N+1)\times2(N+1)$ that is linear in $\omega$ by using \cite{Jansen2017,Mamani:2022akq}
{\small
\begin{equation}
    \hat{N}_0=\begin{pmatrix}
       \hat{M}_0 & \hat{M}_1 \\
        0 & \mathbb{I}
    \end{pmatrix}\ , \ \hat{N}_0=\begin{pmatrix}
       0 & \hat{M}_2 \\
        -\mathbb{I} & 0
    \end{pmatrix}\ , \
 \vec{w}=\begin{pmatrix}
       \vec{a}  \\
        \omega \vec{a}
    \end{pmatrix}\ .
    \label{eq_Ni}
\end{equation}}
Then, a matrix eigenvalue problem is achieved, which can be easily solved through algebraic tools such as \textit{Mathematica}, 
\begin{equation}
    \left( \omega \hat{N}_1+ \hat{N}_0 \right) \vec{w}=0\ .
\end{equation}
The results depend on the algebraic or numerical methods used to solve the specific eigenvalue problem, and spurious eigenvalues may arise. One way to discard these artifical solutions is through the computation of the QNMs by two different grids $N_1$, $N_2$ \cite{Jansen2017}. The actual QNMs must be independent to $N$ up to certain numerical tolerance, so those values that coincide in both grids are chosen as valid within a relative tolerance rate $\epsilon\geq|\omega_2-\omega_1|/|\omega_1|$.
Note that, in some cases, equation \eqref{eq_coefdif} can be even more simplified by using $\phi(x(u),\omega)=e^{i\omega r^*(u)}\psi(x(u),\omega)$. Then, equation \eqref{eq_Hdiff} can be written in the following way,
\begin{align}
   \Big[H(u)^2\partial_{u}^2+\left(H'(u)-2i\omega\right)H(u)\partial_{u} \nonumber\\
   -V^{RW}_l(x(u))\Big]\psi(x(u),\omega)=0\ .
   \label{eq_defhoriz}
\end{align}
Whereas this is not a major simplification, it can still be advantageous in some particular frameworks, such as the case of the presence of an event horizon. The boundary conditions in \eqref{eq_ansatzbc} and \eqref{eq_bcbh} are rewritten as
\begin{equation}
    \psi(x(u),\omega)=u^ch(u), \quad \text{with}\ c=\frac{2i\omega}{H'(0)}\ ,
\end{equation}
while the next steps are kept as described above. 

\subsection{Results}

 \begin{figure*}
    \centering
    \begin{subfigure}{0.66\columnwidth}
        \centering
        \includegraphics[width=\columnwidth]{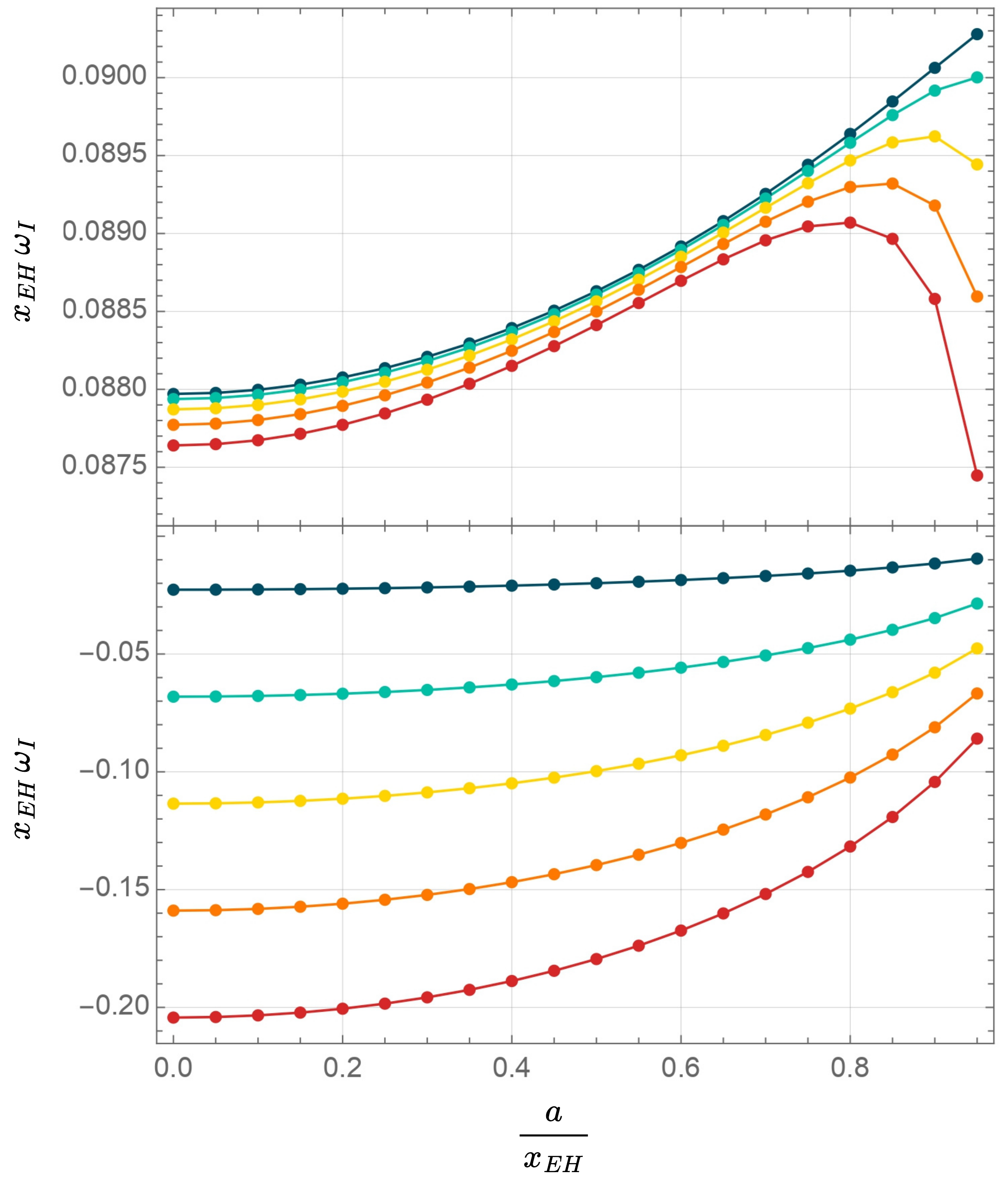}
        \caption{$x_{CH}=1.1\ x_{EH}$}
        \label{fig_qnmhorizon1}
    \end{subfigure}
    \begin{subfigure}{0.66\columnwidth}
        \centering
        \includegraphics[width=\columnwidth]{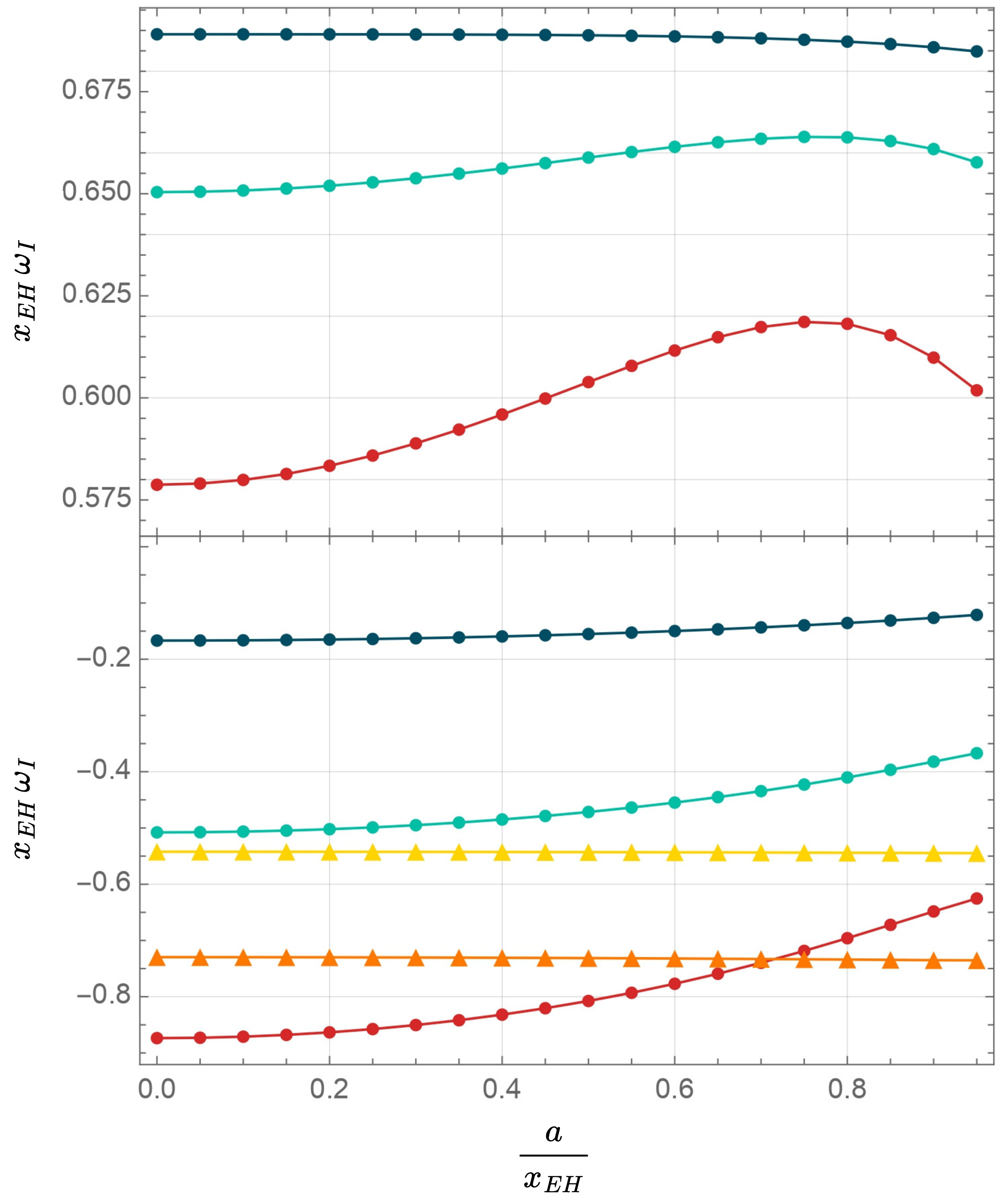}
        \caption{$x_{CH}=5\ x_{EH}$}
        \label{fig_qnmhorizon2}
    \end{subfigure}
    \begin{subfigure}{0.66\columnwidth}
        \centering
        \includegraphics[width=\columnwidth]{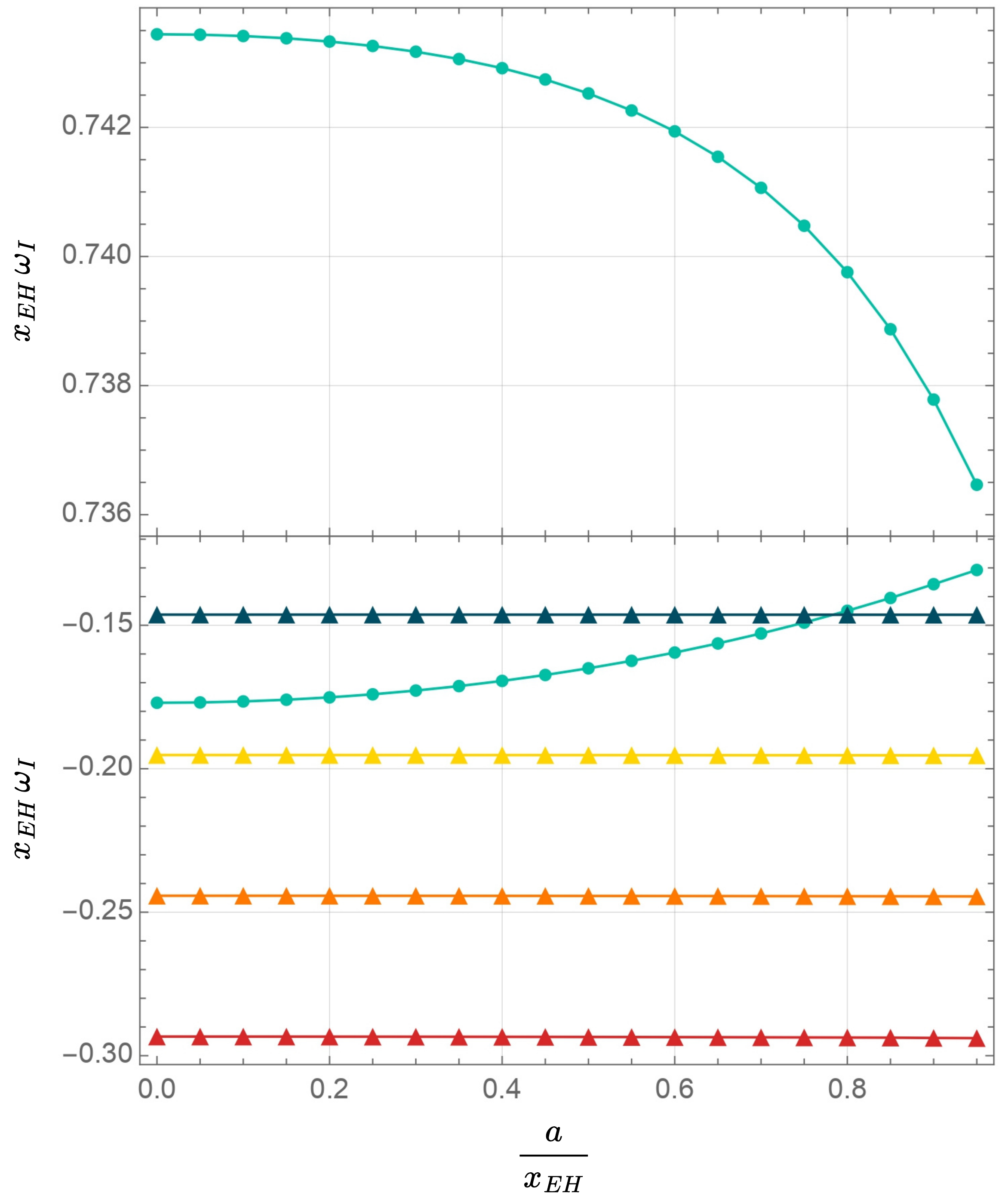}
        \caption{$x_{CH}=20\ x_{EH}$}
        \label{fig_qnmhorizon3}
    \end{subfigure}
    \caption{Fundamental quasinormal mode and first overtones with $l=2$ for an asymptotically de Sitter regular black hole for different values of $x_{CH}$ and $a$. Purely imaginary modes are represented by triangles, while modes with a nonzero real part are represented by dots.}
    \label{fig_qnmhorizon}
\end{figure*}

\begin{figure*}
    \centering
    \begin{subfigure}{0.66\columnwidth}
        \centering
        \includegraphics[width=\columnwidth]{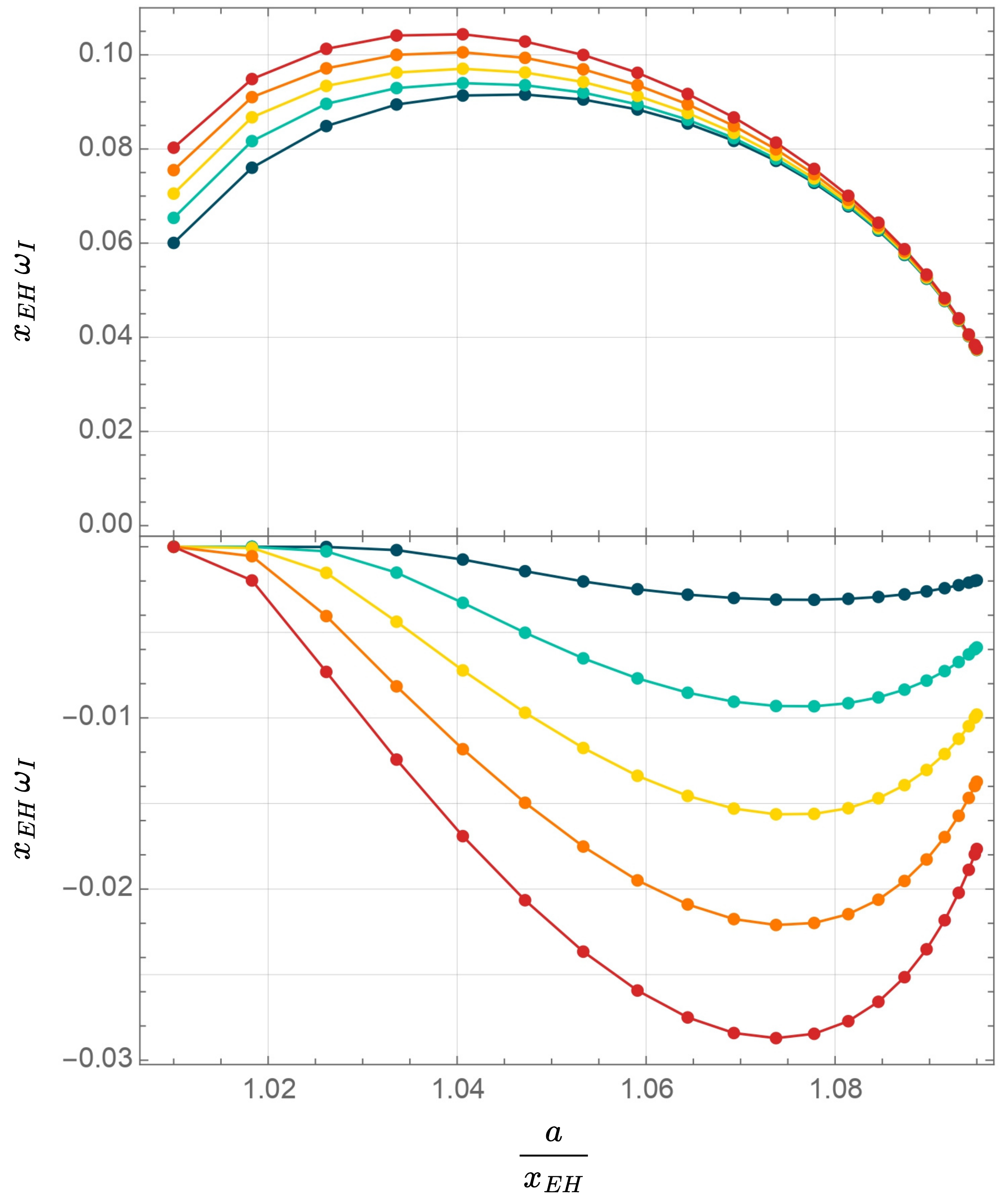}
        \caption{$x_{CH}=1.1\ x_{EH}$}
        \label{fig_qnmhorizonless1}
    \end{subfigure}
    \begin{subfigure}{0.66\columnwidth}
        \centering
        \includegraphics[width=\columnwidth]{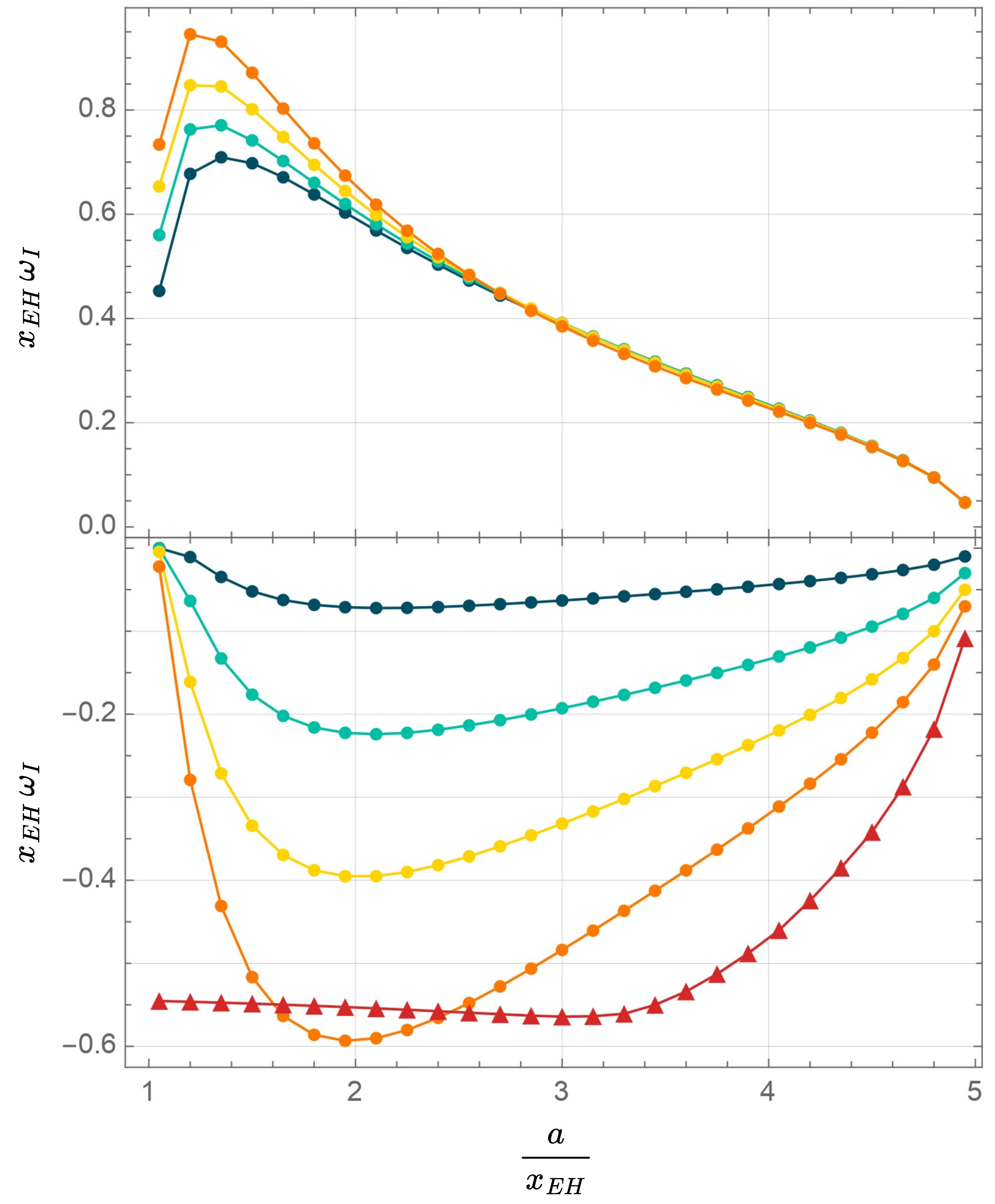}
        \caption{$x_{CH}=5\ x_{EH}$}
        \label{fig_qnmhorizonless2}
    \end{subfigure}
    \begin{subfigure}{0.66\columnwidth}
        \centering
        \includegraphics[width=\columnwidth]{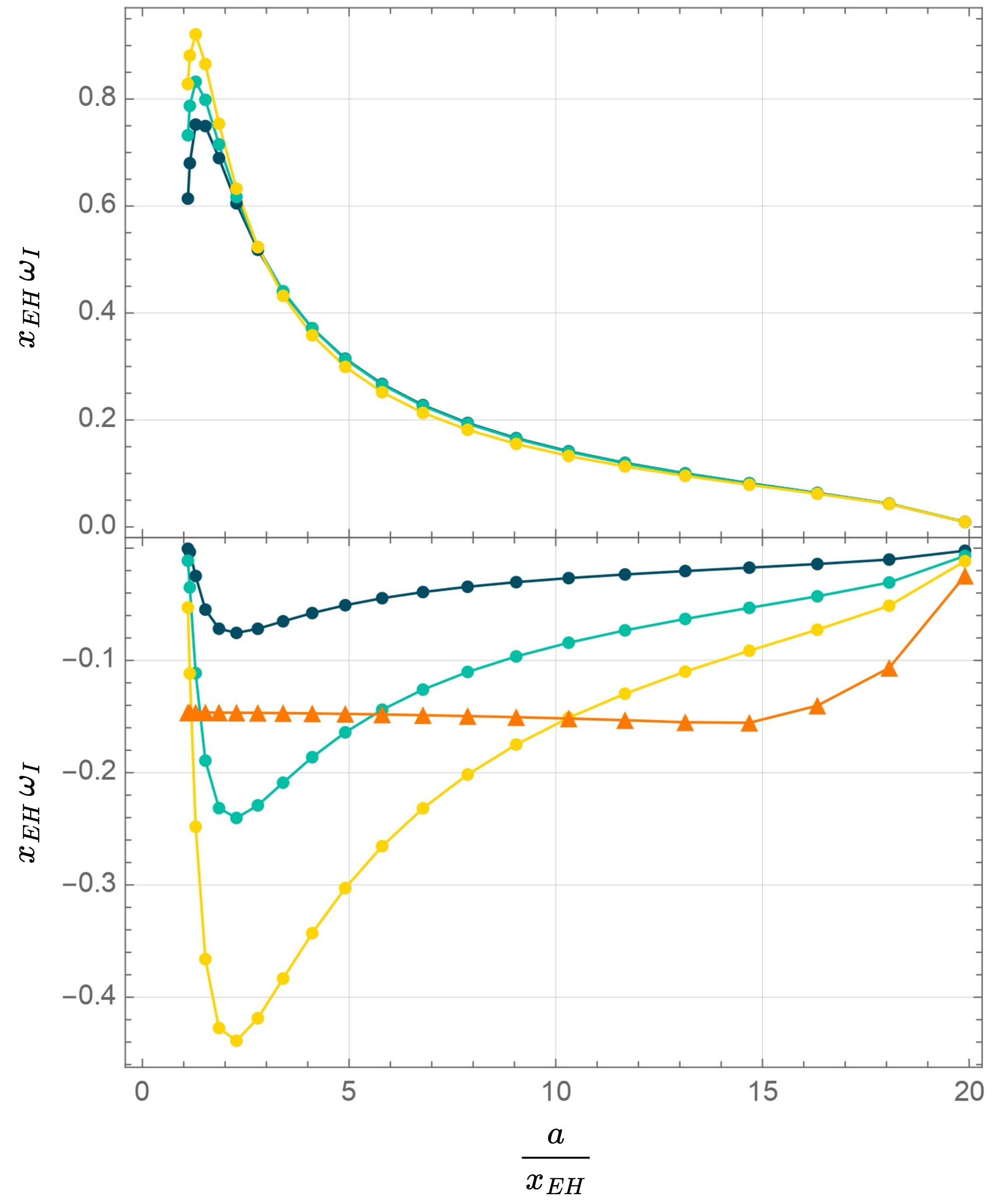}
        \caption{$x_{CH}=20\ x_{EH}$}
        \label{fig_qnmhorizonless3}
    \end{subfigure}
    \caption{Fundamental quasinormal mode and first overtones with $l=2$ for an asymptotically de Sitter traversable wormhole, shown for different values of $x_{CH}$ and $a$. Purely imaginary modes are represented by triangles, while modes with a nonzero real part are represented by dots.}
    \label{fig_qnmhorizonless}
\end{figure*}

To obtain a general idea of how the fundamental QNM and the first overtones change as $a$ increases, three regimes of the Schwarzschild–de Sitter spacetime are considered. Specifically, the near-extremal case with $x_{CH}=1.1\ x_{EH}$, an intermediate case with $x_{CH}=5\ x_{EH}$, and a stonger-de Sitter case with $x_{CH}=20\ x_{EH}$. In every case, $a\in[0,x_{EH})$ for configurations with an event horizon, while $a\in(x_{EH},x_{CH})$ for the horizonless spacetimes.

Similar parameters as in the above section concerning the optical appearance are considered. Note that for $l=2$ there is no qualitative difference between choosing $x_{CH} = 20\, x_{EH}$ or placing it further away. However, the matrices defined in \eqref{eq_Ni} get highly ill-conditioned for high values of $x_{CH}$, so they are avoided.

In general, $N_1=40$, $N_2=42$ and $\epsilon=10^{-3}$ are considered.

\subsubsection{Configurations with an event horizon ($a<x_{EH}$)}

The fundamental QNM and the first few overtones in the case where an event horizon is present have been plotted in Fig. \ref{fig_qnmhorizon}. In both limiting configurations for $x_{CH}$ (when $x_{CH}=1.1\ x_{EH}$ and when $x_{CH}=20\ x_{EH}$), if $a=0$, we have an approximate analytic formula for the corresponding QNMs. Specifically, for a near-extremal Schwarzchild-de Sitter spacetime with $a=0$, the axial gravitational perturbations read \cite{Cardoso2003}
\begin{align}
  \omega_n \sim \kappa_{EH}\Bigg[\pm \sqrt{(l+2)(l-1)-\frac{1}{4}} \nonumber \\
  -i\left(n+\frac{1}{2}\right)\Bigg]\ , \quad n=0,1,2,...\ ,
\end{align}
where $\kappa_{EH}=f'(x_{EH})/2$ is the surface gravity on the event horizon. This is exactly what we observe in Fig. \ref{fig_qnmhorizon1} for $a=0$. The real part depends only on $l$, which is fixed, whereas the imaginary parts depend on the overtone number $n$ and are equally spaced. As $a$ increases, the imaginary parts become closer, while the real parts initially approach each other before diverging again as $a$ approaches $x_{EH}$.

In contrast, for $x_{CH} = 20\ x_{EH}$ and $a = 0$, one expects two different families of QNMs, the purely Schwarzschild QNMs and the purely de Sitter QNMs. The Schwarzschild QNMs cannot be obtained analytically, but it is known from numerical and semi-analytical methods that the fundamental mode is $M\omega_0 = 0.37367 - 0.08896\ i$ \cite{Berti:2009kk}, which in our notation reads $x_{EH} \omega_0 = 0.74912 - 0.17834\ i$. The purely de Sitter QNMs yields \cite{Lopez-Ortega:2006aal}
\begin{equation}
  \omega_n =-i \kappa_{CH}\left(l+n+1\right)\ ,\quad n=0,1,2,..,
\end{equation}
where $\kappa_{CH}=f'(x_{CH})/2$ is the surface gravity at the cosmological horizon. They are purely imaginary and equally spaced for a fixed $l$. Indeed, this is shown  in Fig. \ref{fig_qnmhorizon3}, where all the QNMs are purely imaginary (de Sitter), except for the fundamental mode of Schwarzchild. As $a$ is increased, the purely imaginary modes arise to be unaffected, while the mode with real part behaves as in other regular black holes \cite{Franzin:2023slm}, specifically the Simpson-Visser black bounce. This behavior can be explained by the fact that $a$ modifies the geometry near the event horizon, while the purely imaginary modes are associated with the cosmological horizon, where the geometry remains almost unchanged.

Lastly, for $x_{CH}=5\, x_{EH}$, we observe a transition between the two regimes discussed above. The purely imaginary modes are almost not affected by $a$, while the ones with real part behave similarly to the extremal case ($x_{CH}=1.1\, x_{EH}$) as $a$ increases.

\subsubsection{Configurations without an event horizon ($a>x_{EH}$)}

Firstly, a purely imaginary mode for $x_{CH}= 5\, x_{EH}$ and $x_{CH}= 20\, x_{EH}$ is observed to behave in a different way than other modes. This mode starts at a nonzero value for $a\sim x_{EH}$ and tends to 0 as $a\rightarrow x_{CH}$. Its value gets smaller as $x_{CH}$ increases, gaining relevance on the dynamics, since its damping rate decreases.

\begin{figure}
    \centering
    \includegraphics[width=\linewidth]{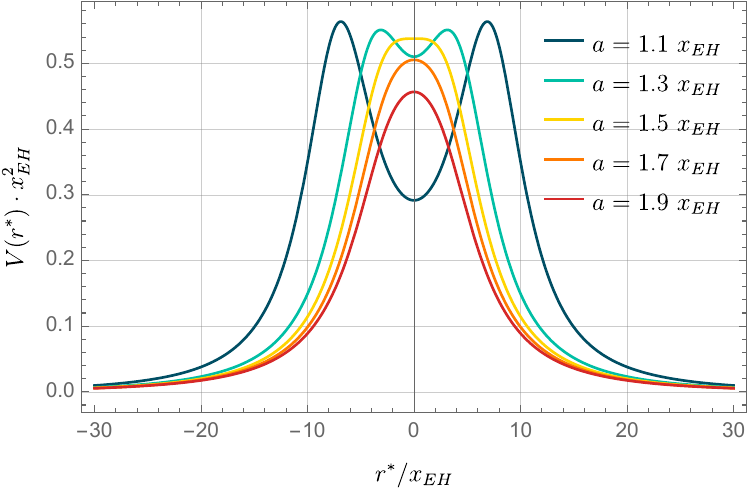}
    \caption{Transition of the multi-peak structure of the potential to a single peak structure for $a\sim a_\text{lim}$. In this case, $x_{CH}=20\, x_{EH}$ and $l=2$ are considered.}
    \label{fig_PotTrans}
\end{figure}

On the other hand, all the computed QNMs with nonzero real part have an imaginary part close to $0$ for $a \sim xb$. This is usual for multi-peak potentials, which arise in some wormhole configurations. Some QNMs get trapped by the potential well generated by the multi-peak structure (see Fig. \ref{fig_PotTrans}) and they become quasi-stationary states, with an imaginary part close to 0. This behavior is observed in all cases ($x_{CH}=1.1\, x_{EH},\, 5\, x_{EH},\, 20\, x_{EH}$). When the well starts getting shallower, fewer quasi-stationary overtones arise, until it disappears at $a_\text{lim}=\frac{3 x_{EH} x_{CH} (x_{EH}+x_{CH})}{2 \left(x_{EH}^2+x_{EH} x_{CH}+x_{CH}^2\right)}$, where there is a single peak at $x=a$ ($r^*=0$). 

Moreover, as $a$ increases, the real parts of the modes become progressively closer until they are nearly identical, while the imaginary parts remain equally spaced. This suggests that the potential could tend to a Posch-teller potential as $a$ increases, since for this potential, the real part of the QNMs depends exclusively on $l$ (so all overtones with fixed $l$ share the same real part), whereas the imaginary part has a linear dependence to $n$, the overtone number (leading to equally spaced modes). 
Let us see how the potential $V(x(r^*))$ behaves for $\frac{|x_{CH}-a|}{|x_{CH}|}<<1$. The tortoise coordinate has a closed analytic form in this regime, given by
\begin{align}
    r^*(x)\sim\frac{1}{2\kappa_{CH}}\log \left(\frac{\sqrt{x_{CH}-a}+\sqrt{x-a}}{\sqrt{x_{CH}-a}-\sqrt{x-a}}\right),
\end{align}
where $\kappa_{CH}$ is the surface gravity of the cosmological horizon and in this limit reads \cite{Griffiths2009}
\begin{equation}
    \kappa_{CH}=\frac{\sqrt{x_{CH}-a} (x_{CH}-x_{EH}) (x_{EH}+2 x_{CH})}{\sqrt{2x_{CH}^3} 
   \left(x_{EH}^2+x_{EH} x_{CH}+x_{CH}^2\right)} .
\end{equation}
The tortoise coordinate is easily invertible in this limit:
\begin{equation}
    x=a+(x_{CH}-a)\tanh^2\left(\kappa_{CH} r^*\right).
    \label{eq_tortoiseposch}
\end{equation}
Since $x\in[a,x_{CH}]$, we define the small quantities $\delta_1=x_{CH}-a$ and $\delta_2=x_{CH}-x$, both assumed to be of the same perturbative order. By expanding the potential as a multivariable Taylor series around $(\delta_1,\delta_2)=(0,0)$ and keeping terms up to linear order, one obtains
{\small
\begin{align}
    V(x)&\sim(x_{CH}-x) \cdot \nonumber\\
    &\frac{\left(l^2+l-2\right) (x_{CH}-x_{EH}) (x_{EH}+2 x_{CH})}{x_{CH}^3 \left(x_{EH}^2+x_{EH}
   x_{CH}+x_{CH}^2\right)}\ .
   \label{eq_potentialposch}
\end{align}}

As a consequence of the terms proportional to $\delta_1\delta_2$ are of second order and $V(x_{CH})=0$, the only remaining term is proportional to $\delta_2$. By substituting \eqref{eq_tortoiseposch} into \eqref{eq_potentialposch}, it leads to
\begin{equation}
    V(r^*)\sim (l+2)(l-1) \frac{V_0}{\cosh^2\left(\kappa_{CH} r^*\right)}\ ,
\end{equation}
where
\begin{equation}
    V_0=(x_{CH}-a)\frac{(x_{CH}-x_{EH}) (x_{EH}+2 x_{CH})}{x_{CH}^3 \left(x_{EH}^2+x_{EH}
   x_{CH}+x_{CH}^2\right)}\ .
\end{equation}
Indeed, this is a Pöschl-Teller potential, as suggested above. This potential has exact QNM solutions \cite{Cardoso2003}
\begin{align}
  \omega_n = \kappa_{CH}\Bigg[\pm \sqrt{\frac{V_0}{\kappa^2_c}(l+2)(l-1)-\frac{1}{4}} \nonumber \\
  -i\left(n+\frac{1}{2}\right)\Bigg]\ , \quad n=0,1,2,...\ ,
\end{align}

\begin{figure}
    \centering
    \includegraphics[width=\linewidth]{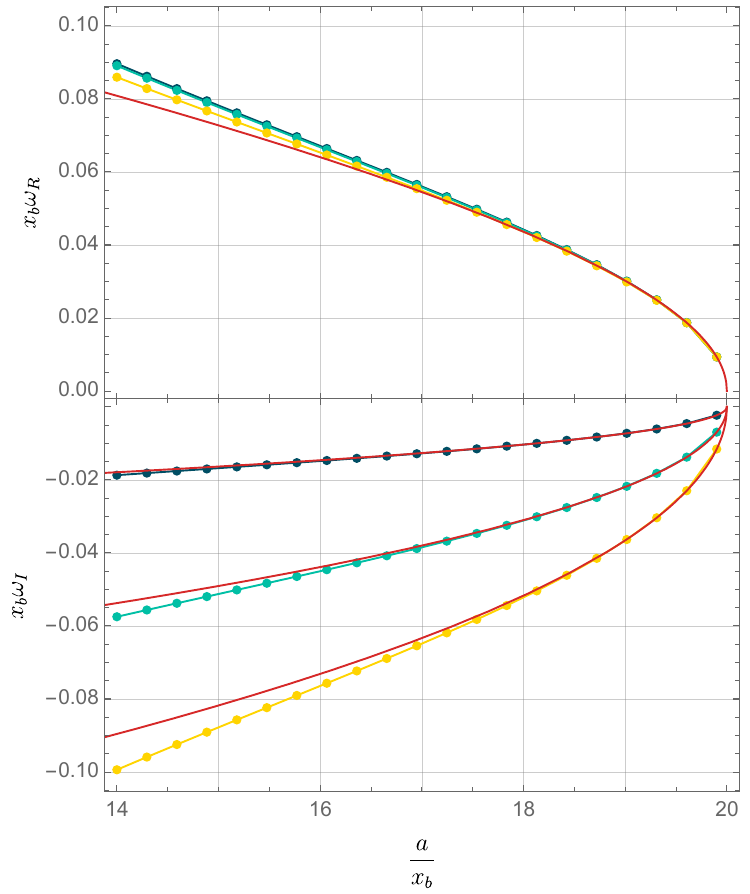}
    \caption{Fundamental quasinormal mode and first overtones of an asymptotically de Sitter traversable wormhole for $x_{CH}=20\ x_{EH}$ and $a\in(14 x_{EH},20x_{EH})$. Numerical modes are plotted with dots, while the analytical approximation is depicted with a straight red line.}
    \label{fig_analitic}
\end{figure}

\begin{figure}
    \centering
    \includegraphics[width=\linewidth]{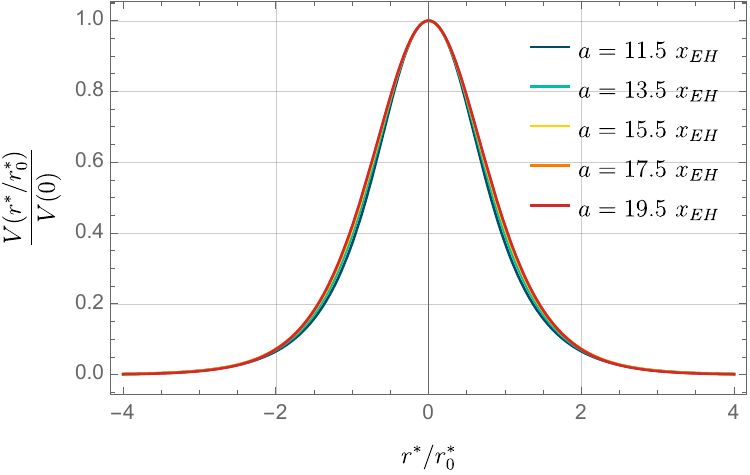}
    \caption{Normalized potential for different values of $a>a_\text{lim}$ for $x_{CH}=20\ x_{EH}$.}
    \label{fig_PT}
\end{figure}

If the analytical approximation is depicted and compared to the numerical results, one gets a quite good correspondence (see Fig. \ref{fig_analitic}). Although the potential becomes exactly Pöschl–Teller only for $a=x_{CH}$, as $a$ moves away from $x_{CH}$, the potential barrier described by $V(x(r^*))$ still has a Pöschl–Teller-like shape. Therefore, the quasinormal modes preserve the behavior despite their values deviate from those predicted in the limit $\frac{|x_{CH}-a|}{|x_{CH}|}<<1$. In fact, if the width of the potential is defined as $r^*_0=\sqrt{-\frac{V''(r^*=0)}{2V(r*=0)}}$, and the height as $V(r*=0)$,the potential shape can be approximated to a Pöschl-Teller potential even when away from the limit $a\sim x_{CH}$, as shown in Fig. \ref{fig_PT}.

\section{Conclusions} \label{Conc}

In this paper, a new class of spacetimes asymptotically de Sitter have been considered which interpolates between a regular black hole and a transversal wormhole. Inspired by the regularization procedure followed in the flat case, where the Schwarzschild metric turns out regular by assuming a finite bounce of the metric \cite{Simpson:2018tsi}, the same procedure is applied to the Schwarzschild-de Sitter spacetime. The result is a regular spacetime asymptotically de Sitter that might own a couple of symmetric event horizons or not, depending on the value of a free new parameter, $a$. One should notice that asymptotically de Sitter spacetimes are inherently interesting since the universe is considered to be roundly de Sitter. Then, the corresponding energy conditions are analyzed showing clear violations, as expected by singularity theorems. In addition, a possible source for this class of spacetimes is reconstructed, which leads to similar results as in the usual case of asymptotically flat black bounces \cite{Alencar:2025jvl}. Indeed, non-linear electrodynamics and a scalar field are necessary to hold the solution of the Einstein field equations. The only novelty is that a cosmological constant is also necessary, as natural due to the asymptotic nature of such a class of spacetimes. Then, two approaches have been followed to point to possible observational signatures of the presence of a cosmological horizon (proper of de Sitter-like spacetimes) and the absence of spacetime singularities. To do so, the optical appearance of such type of objects has been simulated and the spectrum of QNMs have been obtained.\\

 From the side of the images, we have found tiny differences when comparing a singular flat Schwarzschild black hole ($a=0$) with an asymptotic de Sitter Schwarzschild black hole ($a=0$) or with a regular asymptotic de Sitter black hole ($a\neq 0$), as shown in Figs.~\ref{Shadows1} and \ref{Shadows2}. This result points to the same conclusions as previous papers where the regular Simpson-Visser black hole is compared with the Schwarzschild spacetime \cite{Guerrero:2021ues}. However, also as pointed previously, when the free parameter is larger than a particular critical value $a>a_c$, such that the spacetime becomes free of event horizons and a traversable wormhole arises, some differences in the observed luminosity emerge, as shown in Figs.~\ref{Shadows1} and \ref{Shadows2}. Hence, the presence of an event horizon plays a much more important role than the presence (and location) of a cosmological horizon. Indeed, one would expect this, as the photon ring structure of the images is affected mainly by features near the object rather than effects close to the observer. \\

Regarding gravitational-wave emission, the presence of a cosmological horizon as well as the nature of the object (controlled by the parameter $a$) has important effects on the QNMs spectrum. To do so, we have obtained its axial QNMs by using the pseudospectral method. This approach allows to obtain the fundamental QNM and some few overtones for different values of $x_{CH}$ and $a$. In general, we observed two families of QNMs. Those QNMs with a non-zero real part change significantly as $a$ increases whereas purely imaginary modes remain almost unaffected.

For $a=0$ results reproduce exactly the QNMs foun in previous literature for the Schwarzschild-de Sitter spacetime, which allow to verify the consistency of the numerical setup. For $a<x_{EH}$, an event horizon is present, and the QNM behaviour depends strongly on $x_{CH}$, i.e. on the position of the cosmological horizon. In fact, for $x_{CH}>>x_{EH}$, the modes with non-zero real part behave like the Simpson-Visser spacetime QNMs for all $a$ values, while the purely imaginary modes are almost unaffected. This suggests an effective decoupling of the two families of QNMs.

In the horizonless case, if $a\sim x_{EH}$, the QNMs become quasistationary due to the emergence of a potential well that traps these modes. On the other hand, when $a \sim x_{CH}$ the spectrum approaches a Pöschl-Teller potential, which can be verified by showing the potential reduces to this form in this limit. A purely imaginary mode that goes to zero is also observed as $a$ approaches $x_{CH}$. \\

Hence, results from both approaches suggest that while some features might have some effects on some observations, other observational proofs remain unaffected. For instance, the presence (and location) of a cosmological horizon has a direct consequence over the frequencies of the gravitational waves, it does not affect the photon rings substructure in the images around the central object. Then, both observational approaches seem to be complemented for building the basis of a future multi-messenger astronomy. 

\section*{Acknowledgements}

ADC is funded by a pre-doctoral contract from the Predoctoral Contracts UVa 2022 co-financed by Banco Santander. This work is supported by the Spanish National Grant PID2024-157196NB-I00 funded by MICIU/AEI/10.13039/501100011033; the Department of
Education, Junta de Castilla y Le\'on and FEDER Funds, Ref. CLU-2023-1-05.
A.~R. would like to express his gratitude to Silesian University in Opava, Czech Republic, for their financial support.
A.~R. is very grateful for the hospitality of the University of Valencia (Spain), Valencia Polytechnic University (Spain) and
the Complutense University of Madrid (Spain). 
The creation of this article was supported by the grant program Vouchers for Universities in the Moravian-Silesian Region (registration number CZ.10.03.01/00/23\_042/00003901119).
This article is based upon work from COST Action FuSe, CA24101, supported by COST (European Cooperation in Science and Technology).
The authors would like to thank Conselho Nacional de Desenvolvimento Científico e Tecnológico (CNPq), Fundação
Cearense de Apoio ao Desenvolvimento Científico e Tecnológico (FUNCAP) and Coordenação de Aperfeiçoamento
de Pessoal de Nível Superior - Brasil (CAPES) for finantial support.


\clearpage
\appendix

\section{Equations for the QNMs}

In this appendix, we provide the explicit expressions for the nonvanishing functions $c_{ij}(z)$ defined in \eqref{eq_funcionsc}. 
\subsection{Case with a horizon ($a<x_{EH}$)}
In this case we used Eq. \eqref{eq_defhoriz} and obtained the following functions:
\begin{align}
    c_{00}(z)&=-V(u(z)),\\
   c_{01}(z)&=\frac{4 i H(u(z)) }{(z+1)^2 H'(0)}\Big((z+1) H'(u(z))\nonumber \\
   &-2
   H(u(z))\Big),\\
   c_{02}(z)&=\frac{8 H(u(z)) \left((z+1) H'(0)-2 H(u(z))\right)}{(z+1)^2
   H'(0)^2},\\
   c_{10}(z)&=2 H'(u(z)) H(u(z)),\\
   c_{11}(z)&=\frac{16 i H(u(z))^2}{(z+1) H'(0)}-4 i H(u(z)), \\
   c_{20}(z)&=4 H(u(z))^2,
\end{align}
where $u(z)=(z+1)/2$, $H'(u(z))=\partial_u H(u(z))$, and  $H'(0)=\partial_u H(u(z))|_{u=0}$.
\vspace{0.5\baselineskip}
\subsection{Case without a horizon ($a>x_{EH}$)}
In the horizonless regime we used Eq. \eqref{eq_Hdiff} and obtained the following functions:
\begin{align}
    c_{00}(z)&=2 b H(u(z))\frac{ 2 (b+1) H(u(z))-(z-1)
   H'(u(z))}{(z-1)^2}\nonumber\\
   &-V(u(z)),\\
   c_{01}(z)&=\frac{2 i H(u(z))}{(z-1)
   (z+1)^2 H'(0)} \Big(\left(z^2-1\right) H'(u(z))\nonumber \\
   &-2 (2 b (z+1)+z-1)
   H(u(z))\Big),\\
   c_{02}(z)&=1-\frac{4 H(u(z))^2}{(z+1)^2 H'(0)^2},\\
   c_{10}(z)&=2 H(u(z)) \left(H'(u(z))-\frac{4 b
   H(u(z))}{z-1}\right),\\
   c_{11}(z)&=\frac{8 i H(u(z))^2}{(z+1) H'(0)}, \\
   c_{20}(z)&=4 H(u(z))^2,
\end{align}
where $b=0$ for symmetric modes and $b=\frac{1}{2}$ for antisymmetric modes.

\end{document}